\documentclass[5p]{elsarticle}

\usepackage{hyperref}
\usepackage{amsmath,bm}
\usepackage{subcaption}
\usepackage[utf8]{inputenc}

\journal{Journal of Non-Newtonian Fluid Mechanics}

\begin{document}

\begin{frontmatter}

\title{Elastoviscoplastic flows in porous media}

\author[KTH]{F. De Vita\corref{mycorrespondingauthor}}
\cortext[mycorrespondingauthor]{Corresponding author}
\ead{fdv@mech.kth.se}
\author[KTH]{M. E. Rosti}
\author[KTH]{D. Izbassarov}
\author[ENSEEIHT]{L. Duffo}
\author[KTH]{O. Tammisola}
\author[OH]{S. Hormozi}
\author[KTH]{L. Brandt}

\address[KTH]{Linn\'e FLOW Centre and SeRC (Swedish e-Science Research Centre), KTH Mechanics, S-100 44 Stockholm, Sweden}
\address[OH]{Department of Mechanical Engineering, Ohio University, Athens, OH 45701-2979, USA}
\address[ENSEEIHT]{ENSEEIHT, 2, rue Charles Camichel - BP 7122 31071 Toulouse Cedex 7, France}

\begin{abstract}
We investigate the elastoviscoplastic flow through porous media by numerical simulations. We solve the Navier-Stokes equations combined with the elastoviscoplastic model proposed by Saramito for the stress tensor evolution 
\cite{Saramito2007}. In this model, the material behaves as a viscoelastic solid when unyielded, and  as a viscoelastic Oldroyd-B fluid for stresses higher than the yield stress. 
The porous media is made of a symmetric array of cylinders, and we solve the flow in one periodic cell. 
We find that the solution is time-dependent even at low Reynolds numbers as we observe
oscillations in time of the unyielded region especially at high Bingham numbers. 
The volume of the unyielded region slightly decreases with the Reynolds number and strongly increases with the Bingham number; up to $70\%$ of the total volume is unyielded for the highest Bingham numbers considered here. 
The flow is mainly shear dominated in the yielded region, while shear and elongational flow are equally distributed in the unyielded region. We compute the relation between the pressure drop and the flow rate in the porous medium and present an empirical closure as function of the Bingham and Reynolds numbers. The apparent permeability, normalized with the case of Newtonian fluids, is shown to be greater than $1$ at low Bingham numbers, corresponding to lower pressure drops due to the flow elasticity, and smaller than $1$ for high Bingham numbers, indicating larger dissipation in the flow owing to the presence of the yielded regions. Finally we investigate the effect of the Weissenberg number on the distribution of the unyielded regions and on the pressure gradient.
\end{abstract}

\begin{keyword}
Porous media, elastoviscoplastic fluid, Darcy's law
\end{keyword}

\end{frontmatter}


\section{Introduction}
Fluid flows through porous media are relevant for different industrial applications such as oil recovery, polymer extrusion, filtration processes and food processing. They are also present in biological flows that involve mass transfer across organic tissues, like blood vessels, kidney and lungs. Sedimentary rocks and riverbeds are examples of this kind of flows in nature \cite{Dullien2012}. The difficulty in understanding these flows arises from the complex structure of the porous media as well as from its multiscale nature; often, the non-Newtonian behaviour of the involved fluids further complicates the dynamics.

The objective of this work is to investigate the elastoviscoplastic (EVP) flow through a porous medium and understand
the role of yield stress in the relationship between the pressure drop and the flow rate. Hence, the focus of the work will be on non-Newtonian effects in porous media.

\subsection{Porous media}
 
Porous media can be defined as a solid matrix with interconnected cavities distributed inside. Usually, these materials are described from a macroscopic point of view, yet the relevant parameters  strongly depend on the microscopic structure. The characteristic properties of a porous medium are the porosity and the permeability. The porosity, $\varepsilon$, measures the void space inside the material and is defined as the ratio between the volume of the voids and the total volume of the medium, thus, varying between 0 and 1. The permeability, $K$, measures the ability of the flow to pass through the porous medium and has the dimension of a length squared. If the medium is impermeable $K = 0$ whereas, if the medium offers no resistance to a flow, the permeability is infinite. Typical values of porosity for porous media can range from 0.32 for cylindrical packings, to 0.8 for foam metals \cite{Macdonald1979,Beavers1967}. Most synthetic and natural porous media exhibit inhomogeneity due to the randomness of their structure. However, in literature is a common practice to consider as a first approximation the medium as homogeneous and composed of an array of cylinders and consider only a periodic cell \cite{Talwar1995,Liu1998,De2017}, thus we adopted the same discretisation in this study.

From an engineering point of view, it is of primary importance to determine the pressure gradient required to obtain a desired flow rate through the porous medium. The pioneer in the field was Darcy who, in 1856, derived an empirical one-dimensional relation between flow rate and pressure drop, i.e., the well-known Darcy law \cite{Darcy1856}. The law was then extended into three dimensions, see for example \cite{Greenkorn1984}, and also derived theoretically by Whitaker \cite{Whitaker1986}. Note that this analytical relation between the pressure drop and the flow rate across the porous media is valid only for Newtonian fluids. An extension of Darcy law to non-Newtonian flows has been derived using homogenisation theory, see  \cite{Hornung2012}. In particular, Hornung \cite{Hornung2012} states that the system of equations derived with the homogenisation method reduces to a non-linear Darcy-like law only if the flow is directional and that \emph{"We can not always expect to have laws as simple as Darcy's, describing complex natural phenomena"}. It is worth noticing that the law derived in Ref.~\cite{Hornung2012} is based on the assumption of a steady-state flow and in the absence of elasticity effects, \emph{i.e.} generalized Newtonian inelastic fluids. However, many real flows in porous media exhibit both elastic and plastic properties. Additionally, other effects that produce non-Darcy behaviours should be considered, inertia being one example of those; see for example Refs.~\cite{kristlanovitch1940,Roustaei2016}.

\subsection{Non-Newtonian flow in porous media}

The fluids flowing inside porous media exhibit often a non-Newtonian behaviour, characterised by strain- and time-dependent viscosity, yield-stress and/or stress relaxation \cite{pearson2002models}. Many models (constitutive equations) have been proposed in literature to describe these materials, from simple power laws for inelastic shear-dependent viscosity fluids to more complex visco-elastic models like the Oldroyd-B closure. The reader is refereed to Ref.~\cite{Sochi2010} for a detailed review of non-Newtonian flows in porous media.

Viscoelastic fluids exhibit both viscous-fluid and elastic-solid behaviour. Polymer solutions are often modelled as viscoelastic materials. Their behaviour can be described combining the Newton law for viscous fluid with Hooke law for elastic solid, as done originally by Maxwell \cite{Larson1988}. In viscoelastic fluids, the stress depends on the history of the rate of strain, with a fading memory. Hence, for slow deformation viscoelastic materials behave as viscous fluids whereas for fast deformation they behave as elastic solids. For this reason, the important parameter characterising the flow of viscoelastic fluids is the ratio between the material time scale and the time scale of the flow, indicated by the Deborah number (De) {\cite{Poole2012}. 

Flow of viscoelastic materials through porous media has been the object of numerous research in the past owing to the large range of applications involved. Experimental studies have been performed to investigate the pressure drop changes due to elasticity \cite{Marshall1967,james_mclaren_1975,Rodriguez1993} or the instability onset for high Deborah numbers \cite{mckinley_raiford_brown_armstrong_1991,Chmielewski1993,Talwar1995,Pakdel1996}. Numerical simulations have also proved to be a valid tool to investigate viscoelastic flows in porous media, uncovering the details of the fluid deformation and stresses. Alcocer and Singh \cite{Alcocer2002} studied the viscoelastic flow through an array of cylinders using the finite element method and a FENE ({finitely extensible nonlinear elastic) model for the stresses. These authors have shown the influence of the cylinder distribution and Deborah number on the permeability of the medium. Richter et al. \cite{Richter2010} studied the effects of the Reynolds number on the instability of a FENE-P viscoelastic fluid flowing around a cylinder and reported an increase in the drag for higher extensibility of the polymer as well as the suppression of the three-dimensional Newtonian B-mode instability for $Re \sim 300$. A similar study, but for the flow around a sphere is presented in Ref.\ \cite{krishnan2017fully}. The effect of high-Weissenberg number (defined as the ratio between elastic force and viscous force) has been investigated for the Oldroyd-B and Giesekus model in Ref.~\cite{Hulsen2005,Fattal2004}. It is shown that by mean of the log conformation representation it is possible to obtain average convergence of the solution beyond the limiting Weissenberg. More recently, Grilli et al \cite{Grilli2013} have simulated the elastic turbulent  flow of an Oldroyd-B fluid through an array of cylinders. De et al \cite{De2017} have anaylzed the effect of the arrangement of the cylinders on the flow at low Reynolds numbers. They have found differences between symmetric and asymmetric configuration in the flow characterstic and effective viscosity. The effects of elasticity on a flow through a porous medium composed of a random distribution of spheres has been studied in \cite{De2017b, liu2017flow} while the effect of the cylinder distribution on the elastic instability has been investigated in \cite{de2017lane}. Similarly, Shahsavari and McKinley \cite{shahsavari2015mobility} studied the flow of a Carreau fluid through a periodic array of cylinders and proposed a scaling for the mobility functions.

\sloppy 
As already stated before, viscoelasticity is not the only property exhibited by non-Newtonian fluids. Another important characteristic is the yield stress or the ability to sustain shear stress. A yield stress fluid, or visco-plastic fluid, behaves as a solid below the yield stress and as a  fluid above.  
A list of several materials exhibiting yield stress is provided in \cite{Bird1983}. The first attempt to study the rheology of a viscoplastic material dates back to Schwedoff \cite{Schwedoff1900}, based on the Maxwell model. Next, Bingham \cite{Bingham1922} proposed a one-dimensional stress-deformation rate equation that was extended into three dimensions and coupled with the Navier-Stokes equations by Oldroyd \cite{Oldroyd1947}. The latter also proposed a new constitutive equation considering a linear Hookean behaviour before yielding. According to this model, the material is no longer rigid before yielding and the stress is expressed as function of the strain and not the strain rate, leading to a cuspid in the constitutive equation.
 
Experiments of sediments and particles interactions in Carbopol solutions and Laponite suspensions \cite{Putz2008,Gueslin2006, firouznia2018} have shown the loss of symmetry corresponding to the elastic effects. Therefore, it is essential to include elastic effects in dealing with conventional yield stress test fluids such as Carbopol, yet only few works have been conducted to introduce elasticity alongside plasticity. 

Recently, Saramito \cite{Saramito2007} proposed a constitutive equation for  EVP fluid flows based on thermodynamic principles. This model reproduces a viscoelastic solid for stresses lower then the yield stress whereas the material behaves  as a viscoelastic Oldroyd-B fluid for stresses higher then the yield stress. To describe the fluidisation process it uses the von Mises yielding criterion, which has been experimentally confirmed \cite{shaukat2012,Martinie2013}. Other models based on the Papanastasiou regularisation have also been proposed to model yield stress \cite{Park2010,Belblidia2011}, see \cite{Fraggedakis2016} for a detailed analysis of these models.

The model proposed in \cite{Saramito2007} was extended by the same author to account for shear-thinning effects \cite{Saramito2009}. The new model combines the Oldroyd-B viscoelastic model with the Herschel-Bulkley viscoplastic model, a power law with index $n > 0$. When $n = 1$ the model reduces to the one proposed in \cite{Saramito2007}. Benito et al. \cite{Benito2008} derived a minimal, fully tensorial, rheological constitutive equation for EVP flows to study the behaviour of immortal fluids: it can describe a large deformation in the elastic regime and predict viscoelastic fluid after yielding. To avoid the zero loss modulus at low shear, Dimitriou et al \cite{Dimitriou2013,Dimitriou2014} proposed a new constitutive model based on the Isotropic Kinematic Hardening (IKH) idea. This model can predict thixotropic behaviours and has been proven to correctly describe EVP materials such as waxy crude oils. 
Armstrong \emph{et al.} \cite{armstrong2016dynamic} modified the Delaware thixotropic model and proposed a new structure-based model which accounts for shear-induced aggregation. This new model has been shown to correctly describe SAOS (small amplitude oscillatory shear) and weak LAOS (large amplitude oscillatory shear) flows. More recently, Wei \emph{et al.} \cite{wei2018multimode} proposed a new model to predict transient thixotropic EVP (TEVP) flow. They combined the multi-lambda (ML) model of \cite{wei2016quantitative} with the IKH of \cite{Dimitriou2014} deriving a new constitutive equation, called ML-IKH, with 12 parameters. This new model accounts for a non-linear thixotropic kinetic equation, kinematic hardening and linear viscoelasticity and has been proven to correctly describe TEVP flows in many conditions, such as LAOS and shear reversal.

Many attempts have been carried out in recent years to derive experimentally a relation between the pressure drop and the the flow rate in porous media for a yield stress fluid \cite{Al-Fariss1987,Chase2005,Chevalier2013,Chevalier2014}. Balhoff and Thompson \cite{Balhoff2004} noted that it is not clear if a general relation can be found for yield stress fluid flows or it should be derived case-by-case. Cheddadi et al \cite{Cheddadi2011}, in particular, have studied the flow of an EVP fluid around an obstacle and shown that viscous, elastic and plastic effects cannot be considered separately before and after yielding but need to be taken into consideration simultaneously. Only few numerical studies have been conducted for yield-stress and, in particular, for EVP flows, see \cite{Mitsoulis2017} and references therein. Recently, Roustaei et al. \cite{Roustaei2016} have investigated numerically the flow of a yield stress fluid along an uneven fracture. They have shown that the approximation error consequent to the evaluation of the pressure drop using a Darcy-type law strongly depends on the geometry of the problem.

We therefore simulate the EVP flow through a model porous medium, represented by an array of cylinders; here only a symmetric periodic cell is considered. The non-Newtonian flow is simulated by solving the full incompressible Navier-Stokes equations coupled with the model proposed by Saramito \cite{Saramito2007} for the evolution of the EVP stress tensor. In the next section, we present the governing equations and their numerical discretisation, as well as the validation performed to verify our mathematical formulation and its numerical implementation. We describe the chosen flow configuration in section 3 and discuss the results in section 4. In particular, we examine the influence of the Reynolds and Bingham numbers on the flow in the porous media. Finally, we summarise the main findings and conclusions in section 5.

\section{Formulation}

The dynamics of an incompressible EVP flow is fully described by the Navier-Stokes equations with an additional constitutive equation for the non-Newtonian stress tensor. The Navier-Stokes equations expressing the mass conservation and momentum balance read
\begin{subequations}
  \begin{align}
    \nabla \cdot \mathbf{u} &= 0, \label{eqn:mass} \\
    \rho  \frac{D \mathbf{u}}{Dt}  &= -\nabla p +
    \nabla \cdot (2 \mu_f \mathbf{D}) +
    \rho \mathbf{f} + \nabla \cdot \bm{\tau}.
    \label{eqn:momentuum}
  \end{align}
  \label{eqn:NS}
\end{subequations}
In the previous set of equations, the symbol $D/Dt = \partial / \partial t + \mathbf{u}\cdot\nabla$ denotes the material derivative sum of the time derivative and the advection term, $\mathbf{u} = \mathbf{u}(\mathbf{x},t)$ and $p=p(\mathbf{x},t)$ are the velocity and pressure fields, $\rho$ and $\mu_f$ the density and viscosity (assumed to be constant) and $\mathbf{D} = (\nabla \mathbf{u} + \nabla \mathbf{u}^T)/2$ the strain rate tensor. The term $\mathbf{f}$ appearing in the momentum equation is a volume body force used to impose the boundary conditions on the solid boundaries through an Immersed Boundary Method. The body force is applied using the direct forcing of \cite{Fadlun2000} with the volume fraction weighting technique. Finally, the last term $\bm{\tau}=\bm{\tau}(\mathbf{x},t)$ is the EVP stress tensor which accounts for the non-Newtonian behaviour of the fluid. 

\begin{figure}[h]
  \centering
  \includegraphics[width=0.3\textwidth]{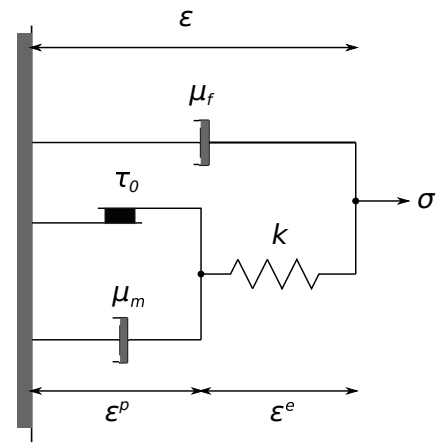}
  \caption{Sketch of the mechanical model of the elastoviscoplastic fluid proposed by Saramito \cite{Saramito2007}.}
  \label{fig:model}
\end{figure}

In the present study, to model the EVP stress tensor, we use the model proposed by Saramito \cite{Saramito2007}. Figure \ref{fig:model} illustrates the underlying idea with a mechanical model: at stresses below the yield stress, the friction element remains rigid, and the whole system predicts only recoverable Kelvin–Voigt viscoelastic deformation due to the spring $\kappa$ and the viscous element $\mu_f$. In this condition, the total stress is given by an elastic and a viscous part $\sigma = \kappa \varepsilon^e + \mu_f \dot{\varepsilon}$. As soon as the stress in the friction element exceeds the yield value $\tau_0$, the element breaks, the additional viscous element $\mu_m$ activates and the deformation of the fluid is that of an Oldroyd-B viscoelastic fluid. As shown in Ref.~\cite{Cheddadi2011}, the total strain rate $\dot{\varepsilon}$ is shared between the elastic contribution $\dot{\varepsilon}^e$ and the plastic contribution $\dot{\varepsilon}^p$. 
It is noteworthy to mention that in EVP models  it is possible to have deformation rate below the yield stress due to the contribution of the elastic deformations.  This feature is absent when dealing with ideal yield stress models. In summary, the model reduces to the Oldroyd-B model for $\tau_0 = 0$, the Bingham model is recovered for $\lambda = 0$, while when both $\tau_0 = 0$ and $\lambda = 0$ the fluid is Newtonian with a total viscosity $\mu$ equal to $\mu_m + \mu_f$.

The  objective frame-independent form of the evolution equation for $\tau$ corresponding to the model previously described can be written as follows
\begin{equation}
  \lambda \overset{\nabla}{\bm{\tau}} +\max \left(0, \frac{|\bm{\tau}_d| - \tau_0}{|\bm{\tau}_d|} \right)\bm{\tau} = 2 \mu_m \mathbf{D},
  \label{eqn:EVP}
\end{equation}
where the above transport equation satisfies the second law of thermodynamics. Here, $\lambda$ is the relaxation time, $\mu_m$ is an additional viscosity, $\tau_0$ the yield stress and $\bm{\tau}_d = \bm{\tau} - 1/N \text{tr}(\bm{\tau})\mathbf{I}$ the deviatoric part of the EVP stress tensor, $N$ being the dimension of the problem. The operator $|\bm{\tau}_d|$ represents the second invariant of the stress tensor. The term $\overset{\nabla}{\bm{\tau}}$ represents the upper convected derivative of the EVP stress tensor defined as \cite{Gordon1972}
\begin{equation}
\overset{\nabla}{\bm{\tau}} = \frac{\partial \bm{\tau}}{ \partial t} + \mathbf{u}\cdot
\nabla \bm{\tau}-\bm{\tau}\cdot \nabla \mathbf{u}-\nabla \mathbf{u}^T\cdot
\bm{\tau}.
\end{equation}
If $U$ is a characteristic velocity of the flow and $L$ is a characteristic length scale, the flow of an EVP fluid through a porous medium can be described by the following non-dimensional numbers: the Reynolds number, expressing the ratio of inertia and viscous forces, $Re = \rho U L / \mu$, the Bingham number, the ratio of the yield stress and viscous stresses, $Bi = \tau_0 L / \mu U$ and the Weissenberg number,  the ratio of the elastic  and the viscous force, $Wi = \lambda U / L$.  

\subsection{Numerical details}
The equations of motion are solved with an extensively validated in-house code 
\cite{picano2015,lashgari2014,rosti_brandt_2017a,rosti_brandt_2017b}.
Equations \eqref{eqn:NS} and \eqref{eqn:EVP} are solved on a staggered uniform grid with velocities located on the cell faces and all the other variables (pressure, stress components and material properties) at the cell centers. Time marching is performed with a fractional-step method \cite{Kim1985} where equations \eqref{eqn:momentuum} and \eqref{eqn:conf} are advanced in time with a third-order Runge-Kutta scheme and a Fast Poisson Solver is used to enforce the divergence-free of the velocity field. All the spatial derivatives are approximated with second-order centered finite differences except for the advection term in equation \eqref{eqn:EVP} where the fifth-order WENO is adopted \cite{Shu2009}.

In order to overcome the well known high Weissenberg number problem, the log conformation method is used to ensure the positive definiteness of the stress tensor \cite{Fattal2005,Hulsen2005,Izbassarov2015}. In this approach, equation \eqref{eqn:EVP} is written in terms of the conformation tensor $\bm{A}$, defined as $\bm{\tau} \lambda / \mu_m + \bm{I}$, and then the logarithm of the conformation tensor $\bm{\Psi}=\log\bm{A}$ is considered in the computations. The core feature of this formulation is the decomposition of the gradient field $\nabla \bm{u}^T$ into two anti-symmetric tensors denoted by $\bm{\Omega}$ and $\bm{N}$, and a symmetric one denoted by $\bm{C}$ which commutes with the conformation tensor, i.e.,
\begin{equation}
 \nabla \bm{u}^T = \bm{\Omega} + \bm{C} + \bm{NA}^{-1}.
 \label{eqn:gradu}
\end{equation}
The final expression of the evolution equation in the variable $\bm{\Psi}$ reads  
\begin{equation}
  \frac{\partial \bm{\Psi}}{\partial t} + \nabla\cdot(\bm{u}\bm{\Psi})
  - (\bm{\Omega} \bm{\Psi} - \bm{\Psi} \bm{\Omega}) - 2\bm{C} =
  \frac{F}{\lambda}(e^{-\bm{\Psi}} - \bm{I}),
  \label{eqn:conf}
\end{equation}
where $F = \max(0,1-\tau_0/|\bm{\tau}_d|)$. Finally, the conformation tensor can be obtained by the inverse transformation as $\textbf{A} = e^{\bm{\Psi}}$.

\subsection{Code validation}
The present implementation for single and multiphase flows of an elastoviscoplastic fluid has been extensively validated in \cite{izbassarov2018computational}, where the details of the algorithm are discussed in further detail. Nonetheless, we report here some classical test cases for viscoelastic and EVP flows for the sake of completeness.

In the first test case, we consider a plane Poiseuille flow with an Oldroyd-B fluid ($Bi=0$). Analytical solutions exist for both the initial transient behaviour and the steady-state solution of the flow \cite{Waters1970}. The simulation is performed in a two-dimensional channel bounded by two parallel walls at distance $h$. The fluid, initially at rest, is accelerated by the application of a constant pressure gradient $dp/dx$ in the axial $x$ direction. We choose the distance between the plates $h$ as the characteristic length scale and the maximum steady-state velocity $U = -h^2dp/dx/8(\mu_f+\mu_m)$ as the velocity scale, and define the following non-dimensional numbers:
\begin{equation*}
  \quad Re=\frac{\rho U h}{\mu_f+\mu_m}, \quad Wi=\frac{\lambda U}{h} \quad
  \text{and} \quad \beta=\frac{\mu_f}{\mu_f+\mu_m}.
\end{equation*}
These are fixed to $Re=0.125$, $Wi=0.125$ and $\beta=0.1$. Numerical simulations are performed on a uniform grid with grid size $\Delta = h/180$, where the no-slip boundary condition is applied at the walls and periodicity in the streamwise direction. The time evolution of the streamwise component of the velocity $u$ at the centerline ($y=h/2$) and of the wall shear stress are depicted in figure \ref{fig:testPoi1}, while the steady state profiles of the streamwise velocity component $u$ and two stress components, $\tau_{xy}$ and  $\tau_{xx}$, are shown in figure \ref{fig:testPoi2}. As can be seen from these figures, there is an excellent agreement between our numerical results and the available analytical solutions, indicating the validity of the numerical implementation.

\begin{figure}[ht]
  \centering
  \begin{subfigure}{0.22\textwidth}
    \includegraphics[width=\textwidth]{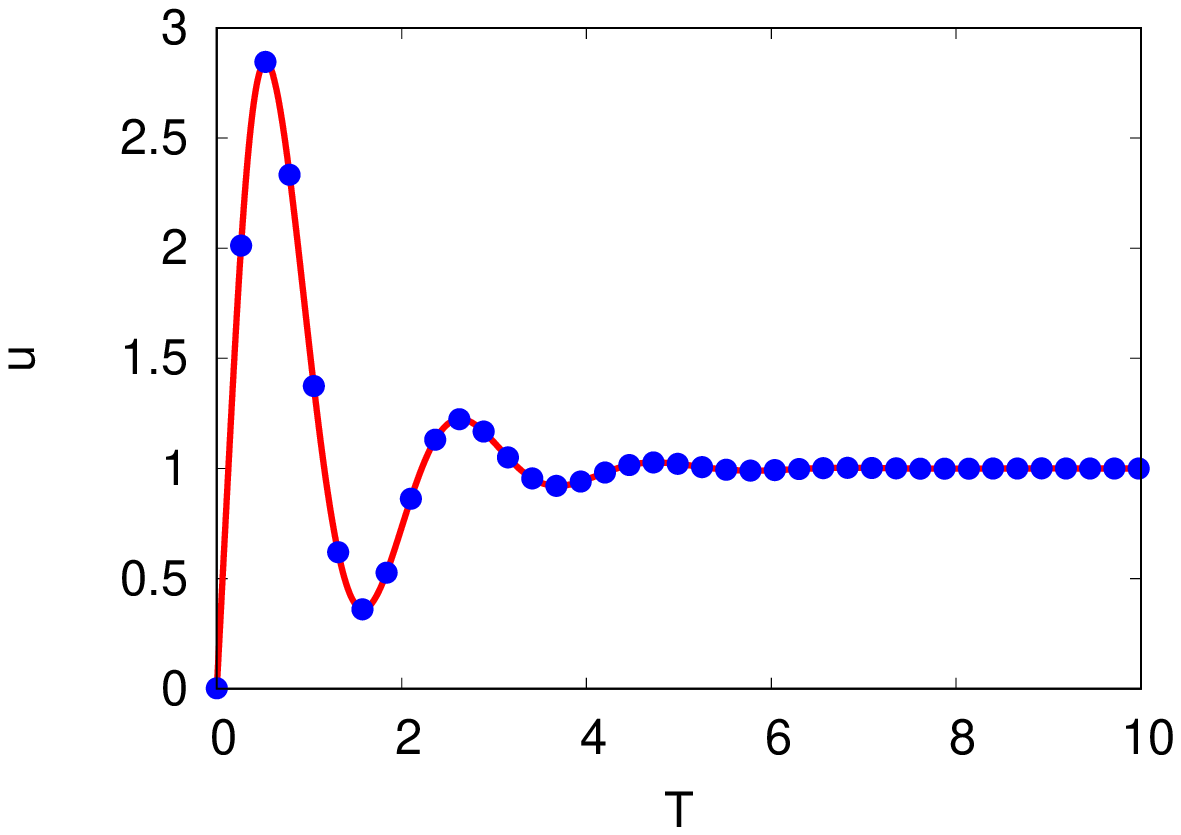}
    \caption{}
    \label{}
  \end{subfigure}
  \begin{subfigure}{0.22\textwidth}
    \includegraphics[width=\textwidth]{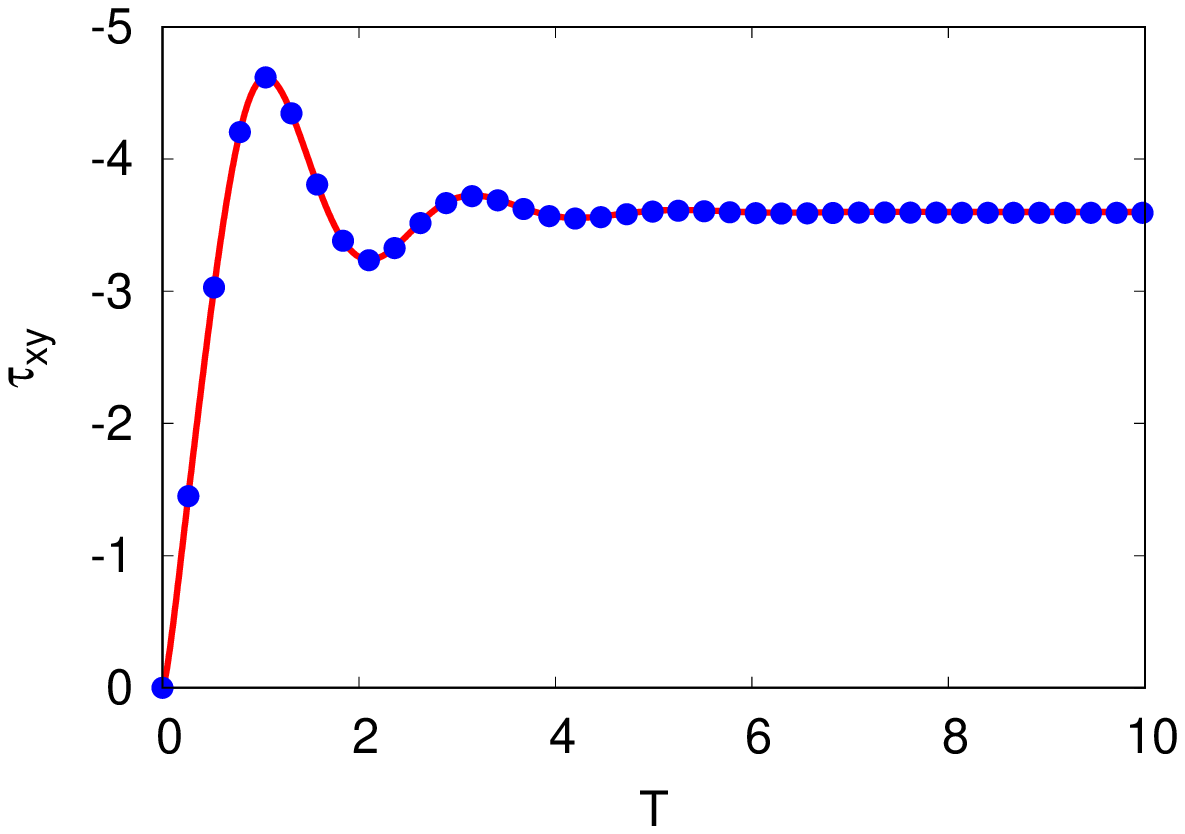}
    \caption{}
    \label{}
  \end{subfigure}
  \caption{Poiseuille flow. Time evolution of (a) the streamwise velocity $u$ at the centerline and  (b) the wall shear stress $\tau_{xy}$. The blue dots represent our numerical results while the red solid lines the analytical solutions in Ref.~\cite{Waters1970}. The time $T$ is made non-dimensional with the viscous time $\rho h^2/(\mu_f + \mu_m)$, and the stress components with $U(\mu_f + \mu_m)/h$.}
  \label{fig:testPoi1}
\end{figure}
\begin{figure}[ht]
  \centering
  \begin{subfigure}{0.22\textwidth}
    \includegraphics[width=\textwidth]{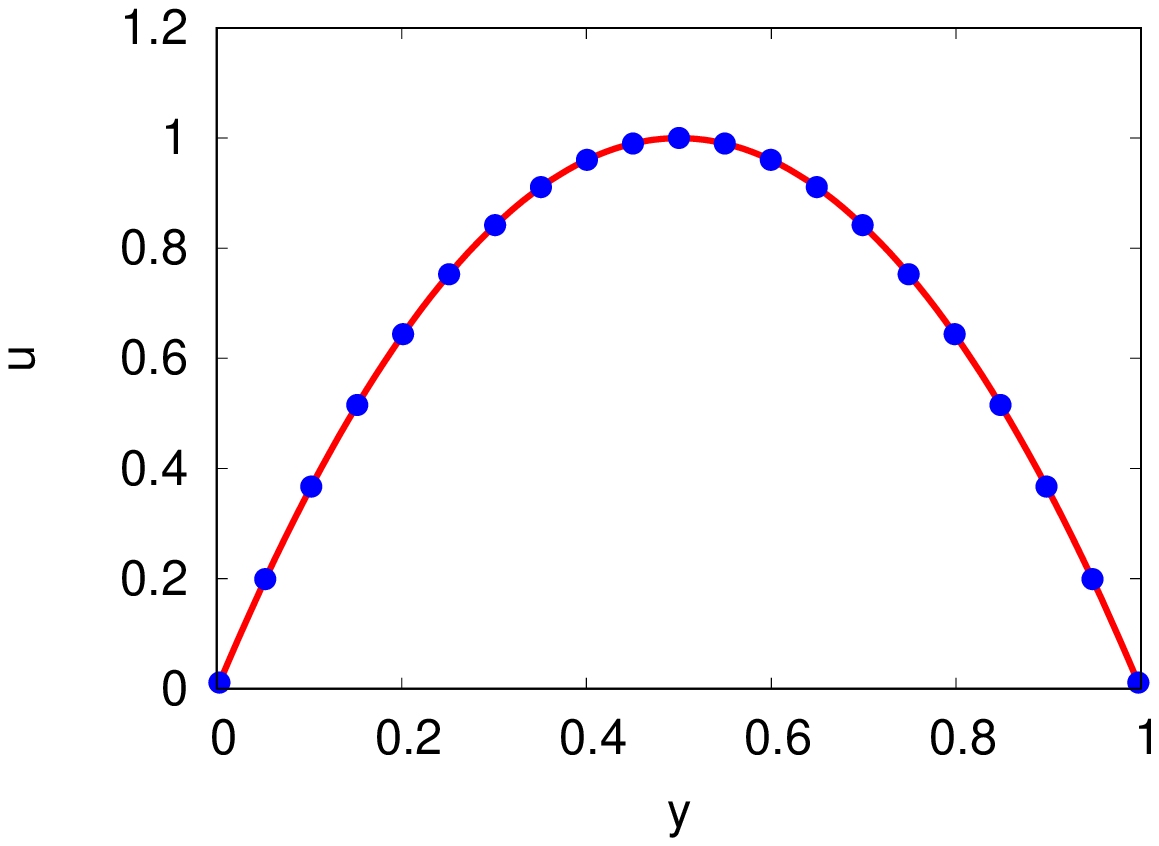}
    \caption{}
    \label{}
  \end{subfigure}
  \begin{subfigure}{0.22\textwidth}
    \includegraphics[width=\textwidth]{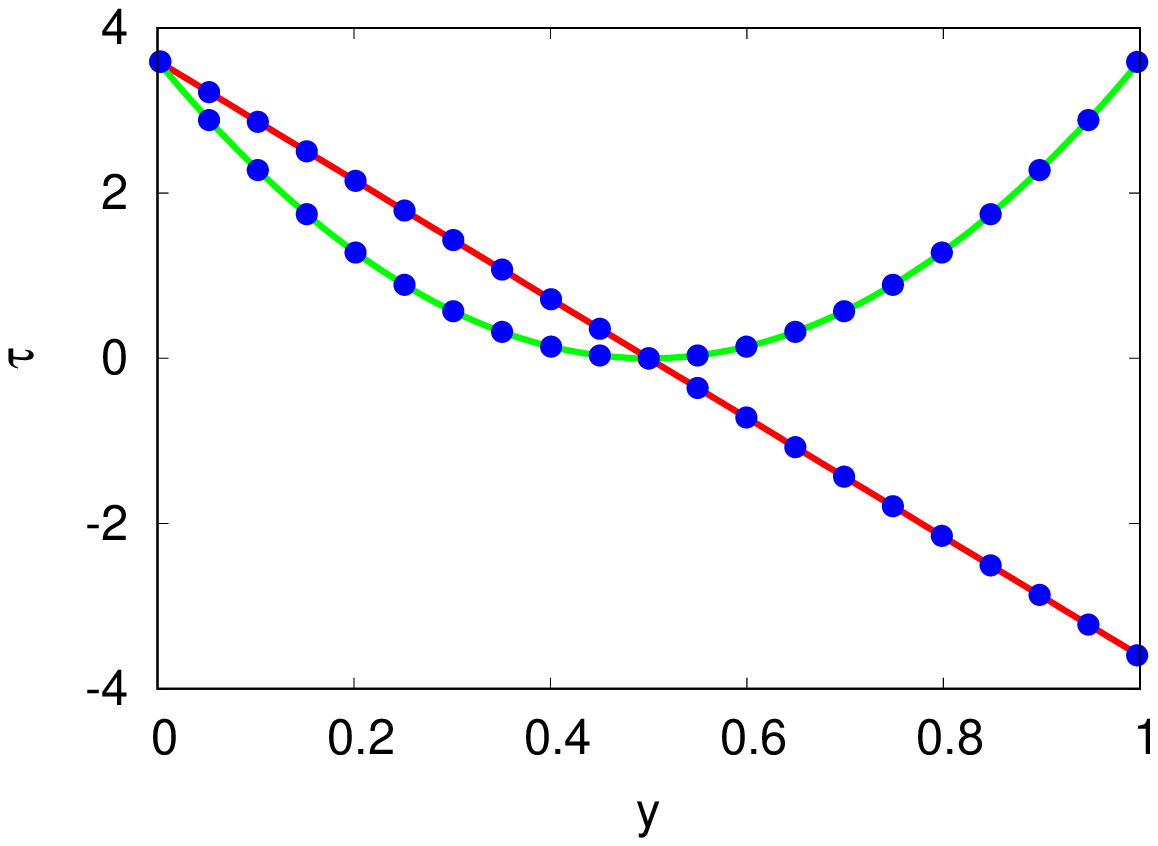}
    \caption{}
    \label{}
  \end{subfigure}  
  \caption{Poiseuille flow. Steady state profiles of (a)  the streamwise velocity component $u$, and (b)  the two stress components, $\tau_{xx}$ (green line) and $\tau_{xy}$ (red line). The blue dots represent our numerical results while the solid lines the analytical solutions in Ref.~\cite{Waters1970}. The stress components are made non-dimensional with $U(\mu_f + \mu_m)/h$, as in figure \ref{fig:testPoi1}.}
  \label{fig:testPoi2}
\end{figure}

Next, we validate the EVP fluid implementation by considering two cases. The first test case is a simple shear flow, with the flow driven by a constant shear rate $\dot{\gamma}_0$. The fluid flow is assumed to have a constant velocity gradient $\nabla {\bf u}=\bigl[\begin{smallmatrix} 0&1 \\ 0&0 \end{smallmatrix} \bigr] \dot{\gamma}_0$, while the Weissenberg number $Wi=\lambda \dot{\gamma}_0$ and the Bingham number $Bi=\tau_0/(\mu_0 \dot{\gamma})$ are fixed to 1 and $\beta=1/9$. The time evolution of the non-zero EVP stress components ($\tau_{xx}$, $\tau_{yy}$ and $\tau_{xy}$) is displayed in figure \ref{fig:shear-a}; the comparison with the results by Saramito \cite{Saramito2007} shows again good agreement.

Finally, we consider the periodic shear flow of an EVP fluid. An oscillatory flow is created by imposing an oscillatory uniform shear strain $\gamma_0 sin(\omega t)$,  where $\gamma_0$ is the strain amplitude and $\omega$ the angular frequency of the oscillations. The Weissenberg number is defined as $Wi=\lambda \omega$ and the Bingham number as $Bi= \tau_y/(\rho \gamma_0 \omega)$; the former is kept constant at $Wi=0.1$, while two values of the latter are considered, $Bi=0$ and $300$. Note that, for $Bi=0$ the material behaves like a viscoelastic fluid (Oldroyd-B) while for $Bi=300$ as an elastic solid. $\beta$ is assumed null in these simulations, i.e., $\mu_f=0$. The evolution of $\tau_{xy}$ is plotted in figure \ref{fig:shear-b} for the two cases considered and compared with the analytical solution provided by Saramito \cite{Saramito2007}. Good agreement is found between the results from our simulation and the analytical solutions also in this configuration.

\begin{figure}[ht]
  \centering
  \begin{subfigure}{0.22\textwidth}
    \includegraphics[width=\textwidth]{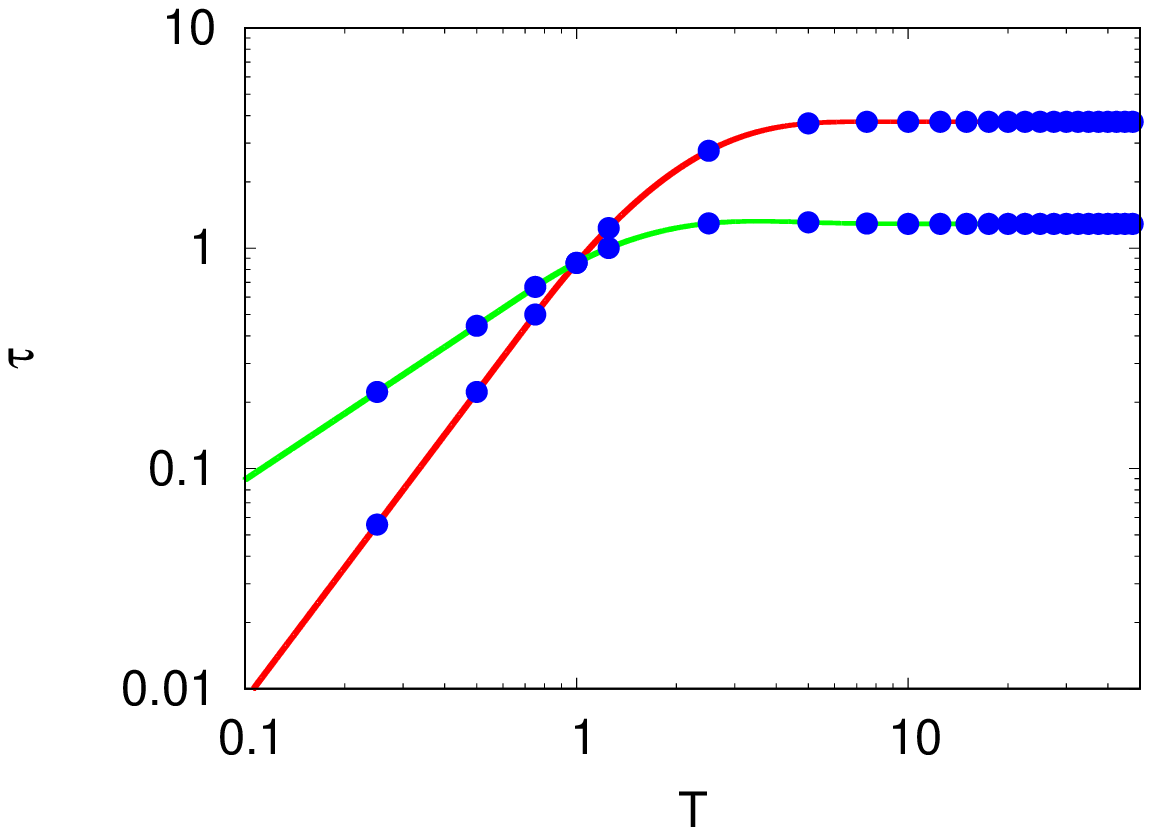}
    \caption{}
    \label{fig:shear-a}
  \end{subfigure}
  \begin{subfigure}{0.22\textwidth}
    \includegraphics[width=\textwidth]{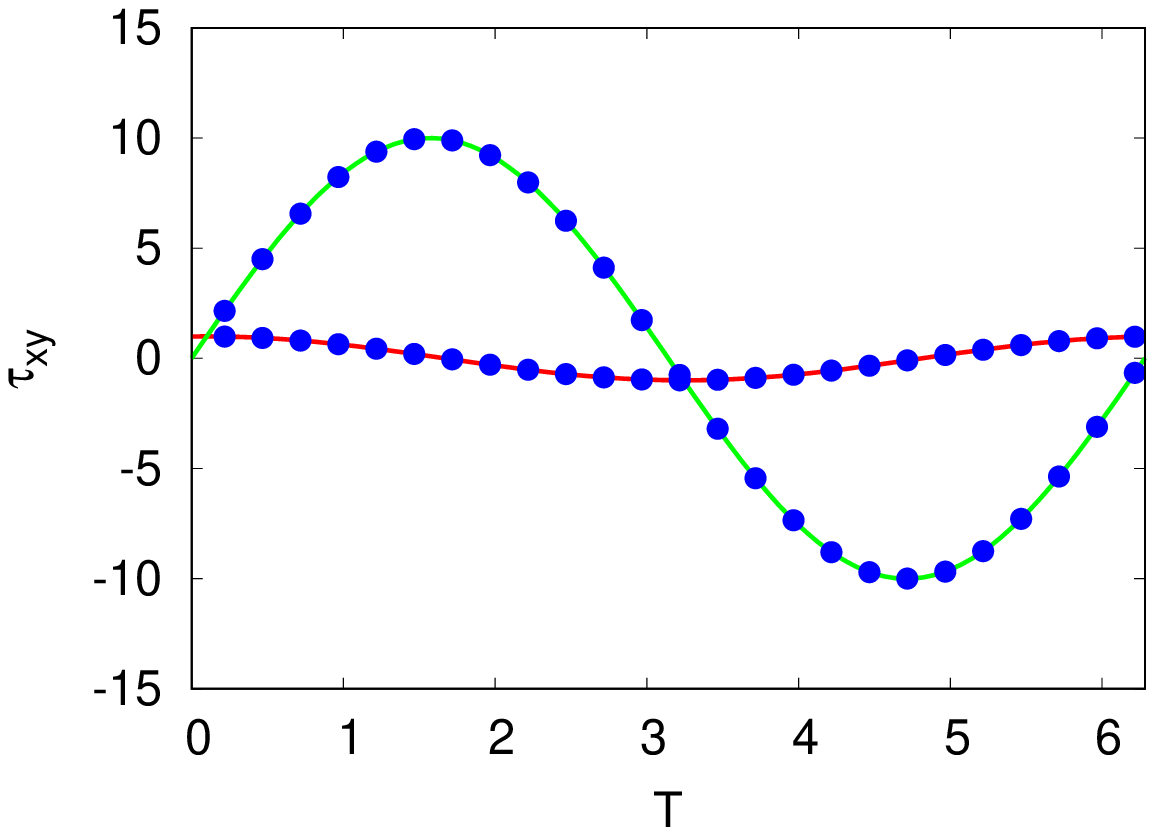}
    \caption{}
    \label{fig:shear-b}
  \end{subfigure}
  \caption{Stationary and oscillating shear flow. (a) The evolution of $\tau_{xx}-\tau_{yy}$ (red line) and $\tau_{xy}$ (green line) in a stationary shear flow. (b) The evolution of the shear stress $\tau_{xy}$ in an oscillating shear flow at $Bi=0$ (red line) and $Bi=300$ (green line). The solid lines represent the analytical solution in Ref.~\cite{Saramito2007} while the blue dots indicate our numerical results. The stress components in the left panel are made non-dimensional with $(\mu_f + \mu_m) \dot{\gamma}_0$, while in the right panel with $(\mu_f + \mu_m) {\gamma}_0 \omega$.}
  \label{fig:testshear}
\end{figure}

\section{Problem description}

We consider the incompressible EVP flow through a model porous medium composed by an array of cylinders with porosity $\varepsilon=0.38$. Here, we focus on a single periodic cell, as sketched in figure \ref{fig:domain}, following the work by \cite{De2017}. 

\begin{figure}[h]
  \centering
  \includegraphics[width=0.3\textwidth]{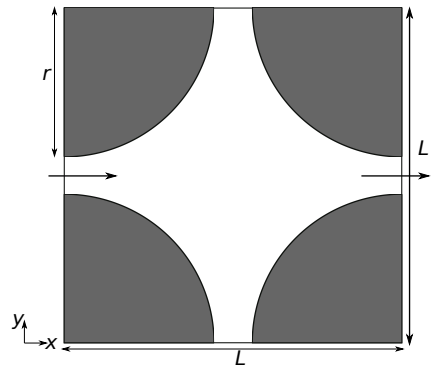}
  \caption{Sketch of the computational domain.}
  \label{fig:domain}
\end{figure}

The numerical domain is a square box of size $L = 2.25r$, $r$ being the radius of the cylinders centered in each corner of the domain and the length-scale of the problem. A periodic boundary condition is enforced in the streamwise $x-$direction, while the free-slip boundary condition is enforced in the $y-$direction, to properly reproduce the effects of adjacent cells of the porous media. Note that, no-slip boundary conditions in the $y-$direction have been tested, providing only slight changes to the flow. Finally, the no-slip boundary condition is enforced on the surface of the cylinders.

We initialize the flow by setting the velocity field and the EVP stress tensor to zero at the beginning of the simulation. The flow is driven at a constant flow rate so that the bulk Reynolds number is $Re = \rho U r / \mu$, where the characteristic velocity $U$ is the bulk flow velocity defined as the volume average of the horizontal velocity $u$, and $\mu = \mu_f + \mu_m$ is the total viscosity and is kept constant. Thus, we compute the streamwise pressure gradient required to provide the desired flow rate at each timestep. To investigate the effects of the flow rate, the role of weak inertial effects, and of the yield stress on the flow in the porous medium, we perform a series of simulations with different Reynolds numbers (0.1, 0.4, 0.8, 1.6) and different Bingham numbers (0, 0.1, 1, 10 and 100), the latter defined as $Bi = \tau_0 r / \mu U$ . Whenever not specified, in all our simulations we fix the Weissenberg number, defined as $Wi = \lambda U / r$, to a constant intermediate value $Wi = 0.5$ and the viscosity ratio $\beta = \mu_f / \mu =0.5$, chosen as in De et al.~in \cite{De2017}. In this way, the viscoelastic behavior of the flow does not change, and we focus uniquely on the role of the Reynolds or the Bingham number. We performed preliminary simulations with a grid size of $L/128$, $L/256$ and $L/384$ and found no significative differences for the last two configurations. For this reason, in all the simulations, we use a uniform constant grid size equal to $L/256$ in all the directions. Note again that the flow geometry and some of the parameters used are the same as De et al.~in \cite{De2017} who studied the viscoelastic flow in a model porous media with the FENE-P model.

\section{Results}

First, we provide a qualitative picture of the flow through the porous medium considered hereafter.
Figure \ref{fig:Newtoniancomparison} shows streamlines and contours of the streamwise velocity component at a Reynolds number of $0.01$ for a Newtonian fluid. The flow enters the domain in the narrow neck between two adjacent cylinders, strongly decelerates due to the sudden expansion of the geometry and then accelerates again due to the contraction formed by the following series of cylinders. Close to the top and bottom boundary the flow generates slowly counter-rotating vortices. We find good agreement between the results of this study and those in literature (see for example figure 3 in \cite{De2017}).

\begin{figure}[ht]
  \centering
  \includegraphics[width=0.25\textwidth]{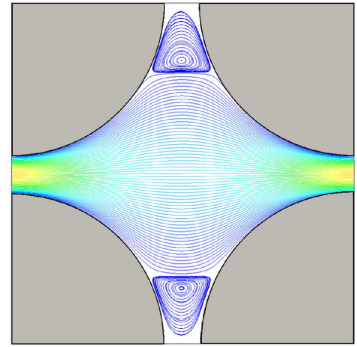}
  \caption{Streamlines, colored by the streamwise velocity component normalized with the maximum velocity (red is the maximum equal to 1, blue is the minimum equal to 0).}
  \label{fig:Newtoniancomparison}
\end{figure}

Next, we investigate the EVP flow through the same porous media. Despite the low Reynolds numbers investigated, we find the solution to be unsteady and not periodic in time; especially for the highest Bingham numbers under investigation, this leads to time oscillations of the pressure drop and of the yielded/unyielded regions; thus a full statistical analysis (e.g., mean values and fluctuations) is required to fully describe the flow. 
It is worth noting that a similar behaviour has been reported previously for viscoelastic flows \cite{De2017,Sousa2010, Hormozi2011, Hormozi2012, Hormozi2014}. The fact that the  steady state flow is not established implies that the flow is not stable to finite spatial perturbations, i.e, cylindrical obstacles. Therefore, the stability  analysis of these EVP flows deserve further investigations. 

Figure \ref{fig:solid_time} shows the evolution of the unyielded region at three selected time instants for the case with $Re=0.1$ and $Bi=100$, one of those displaying the largest variations in time. The regions where the fluid is not yielded are mainly located at the center of the domain and in the two cavities on the vertical centerline; the size of the unyielded region in the centre changes over time, growing in size in the vertical direction and merging with the solid-like region in the narrow gaps, alternatively with the regions on the top and bottom. The central unyielded block also stretches in the streamwise direction, mainly in correspondence of its tail. We also note the presence of thin \textit{fingers} repeatedly appearing and disappearing between the separate unyielded regions. As a consequence of this unsteady behaviour, there can be an instantaneous loss of symmetry with respect to the horizontal axis.

\begin{figure}[ht]
  \centering
  \begin{subfigure}{0.15\textwidth}
    \includegraphics[width=\textwidth]{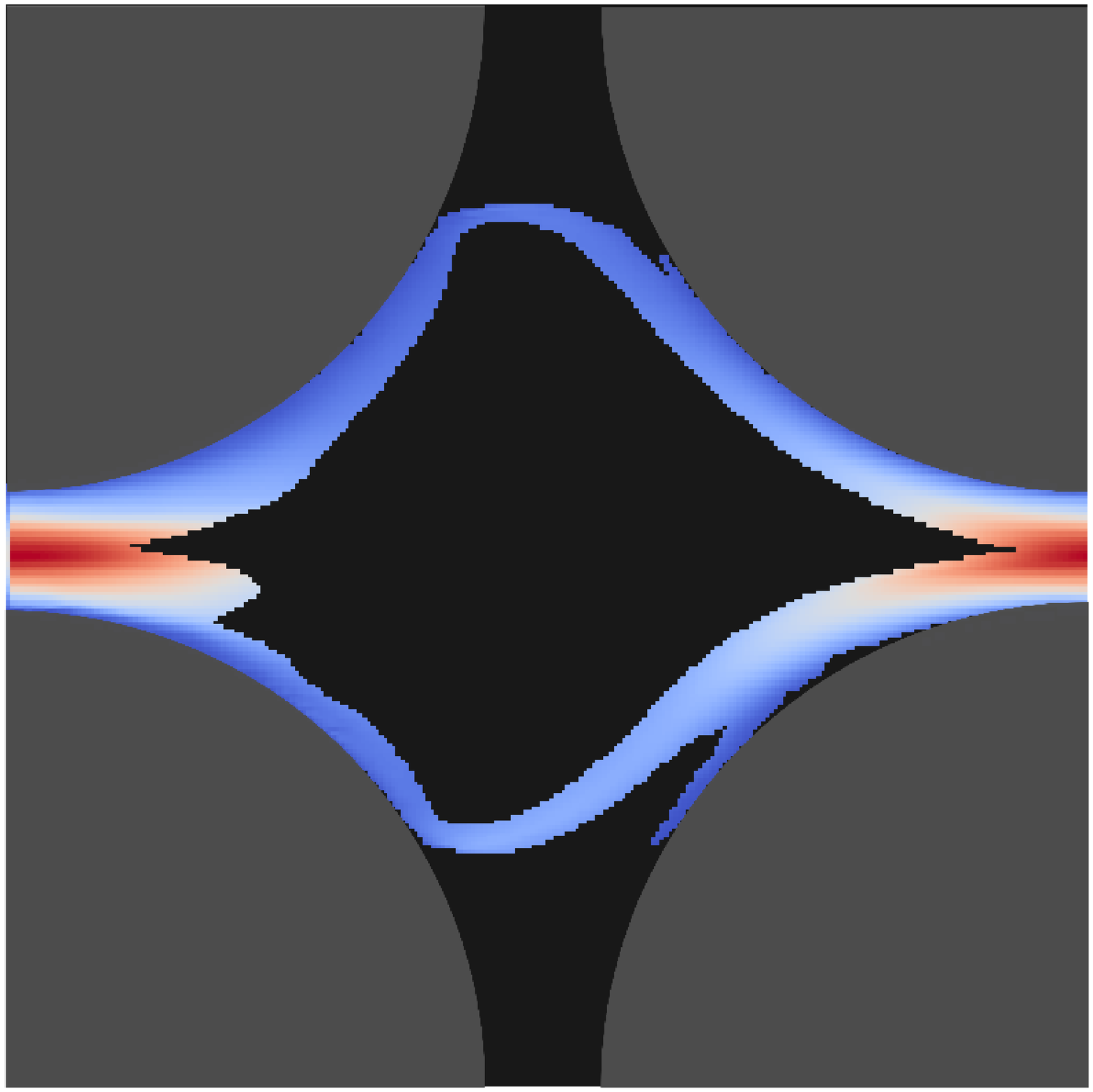}
    \caption{}
    \label{fig:33}
  \end{subfigure}
  \begin{subfigure}{0.15\textwidth}
    \includegraphics[width=\textwidth]{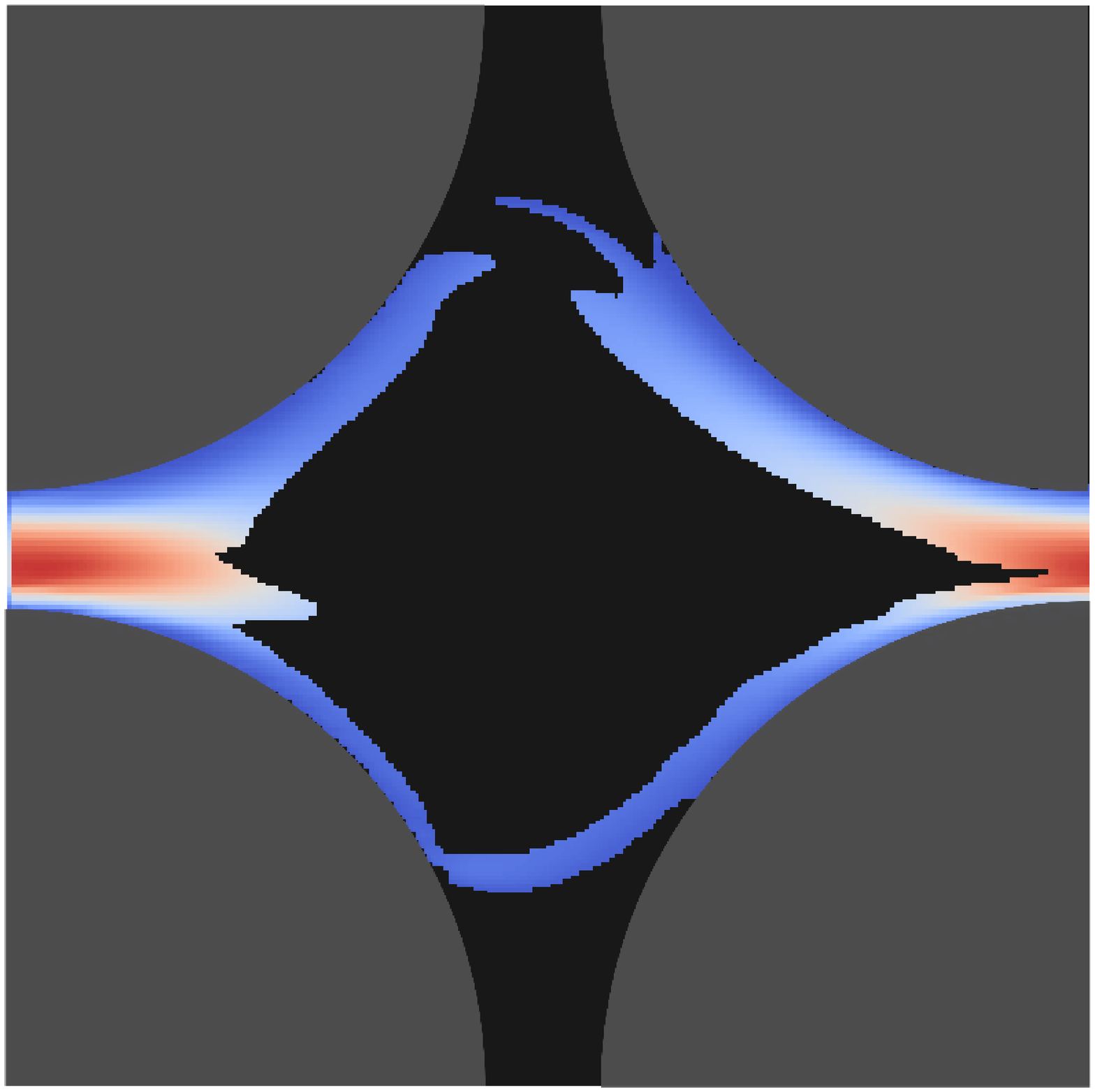}
    \caption{}
    \label{fig:147}
  \end{subfigure}
  \begin{subfigure}{0.15\textwidth}
    \includegraphics[width=\textwidth]{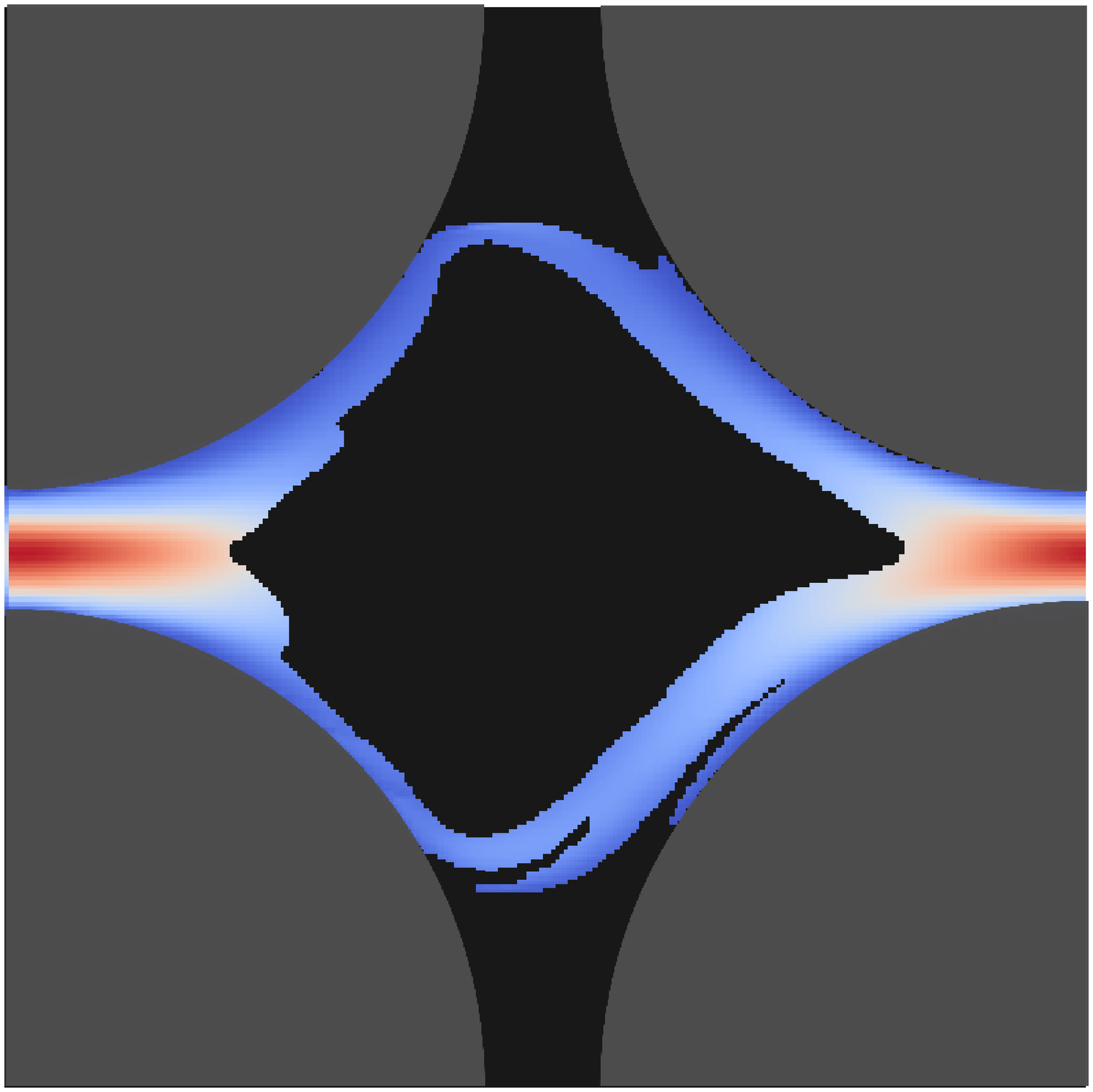}
    \caption{}
    \label{fig:348}
  \end{subfigure}
      \caption{Time evolution of the unyielded region, colored in black, for the EVP fluid in the model porous medium at $Re = 0.1$ and $Bi = 100$. In the figures we also show the contours of the streamwise velocity component: red is used for the maximum velocity (5.6) and blue for the minimum (-0.12). The snapshots are taken at time: a) t = 6.38, b) t = 7.1 and c) t = 8.34. Time is made non-dimensional with  $r/U$.}
  \label{fig:solid_time}
\end{figure}

\begin{figure*}[h]
  \centering
  \begin{subfigure}{0.24\textwidth}
    \includegraphics[width=\textwidth]{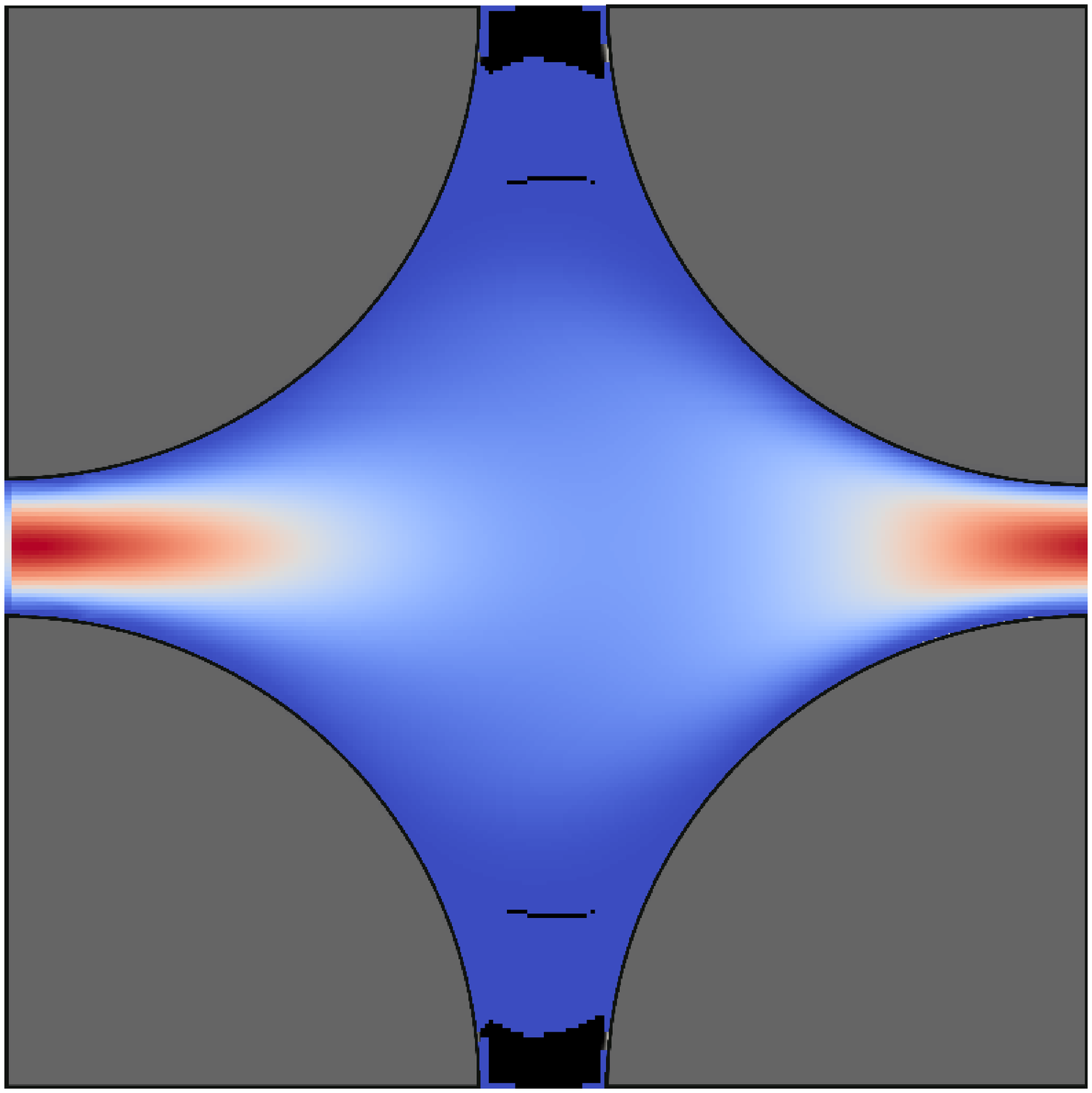}
    \caption{}
    \label{fig:sR_4_01}
  \end{subfigure}
  \begin{subfigure}{0.24\textwidth}
    \includegraphics[width=\textwidth]{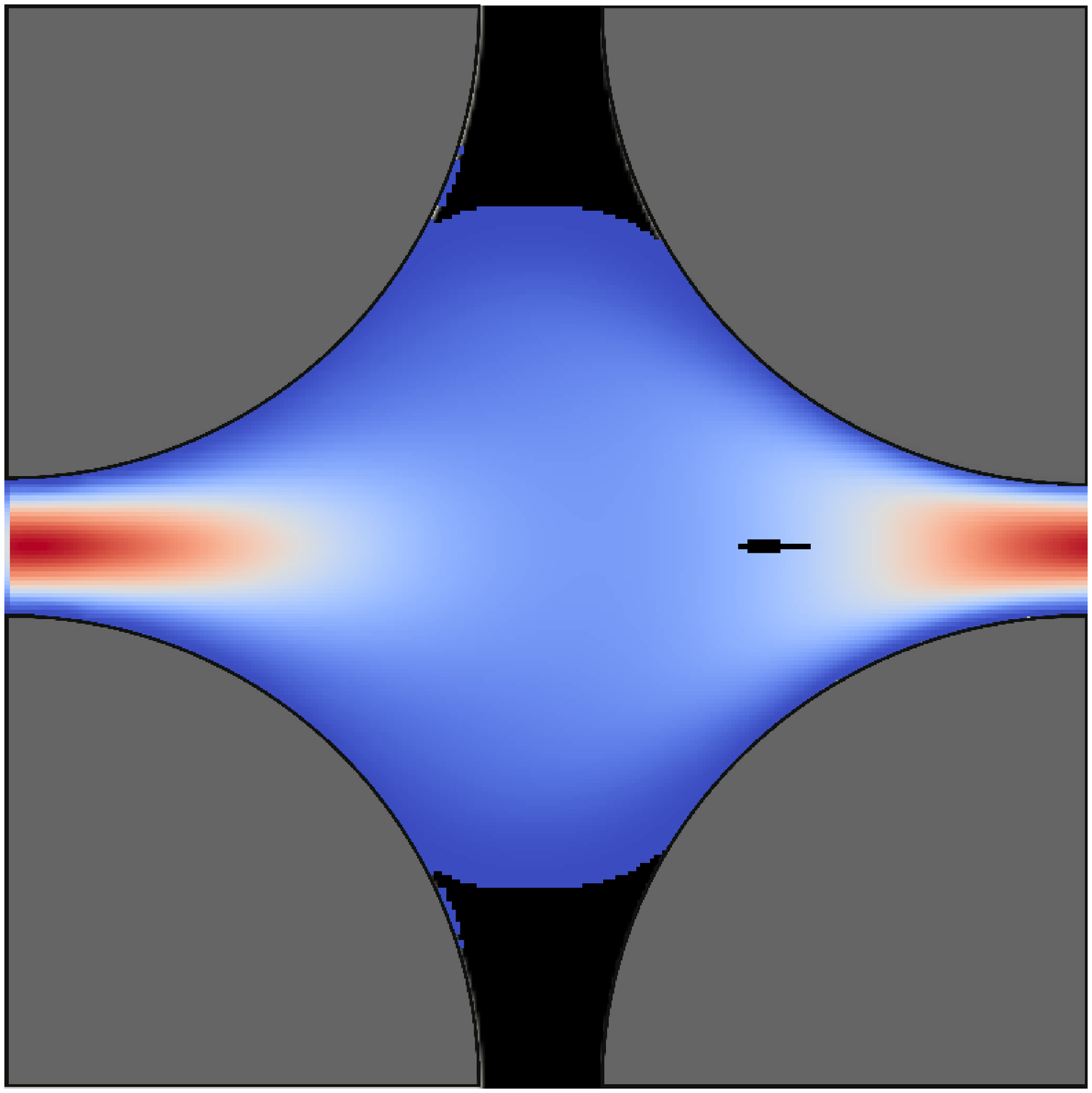}
    \caption{}
    \label{fig:sR_4_1}
  \end{subfigure}
  \begin{subfigure}{0.24\textwidth}
    \includegraphics[width=\textwidth]{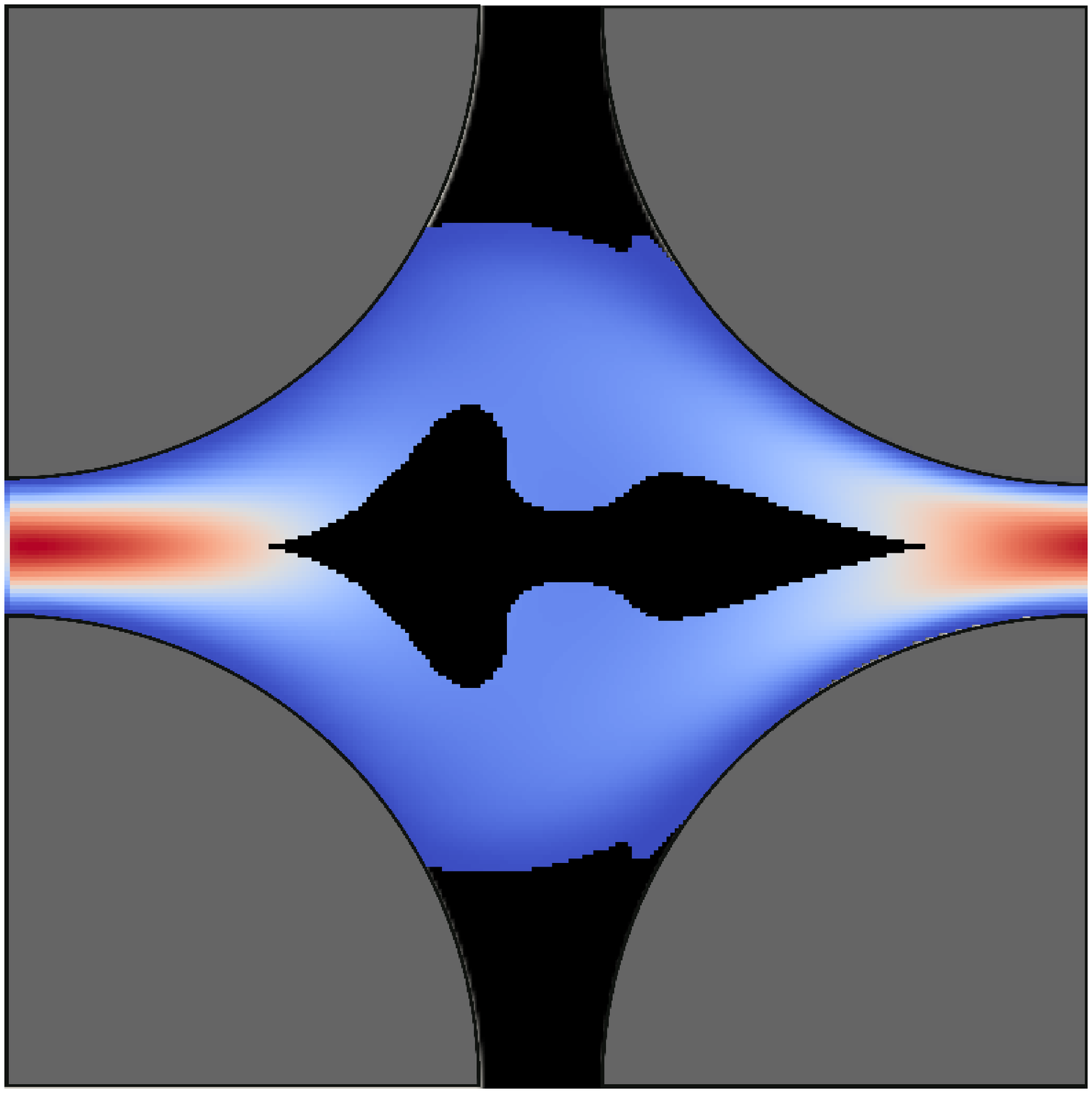}
    \caption{}
    \label{fig:sR_4_10}
  \end{subfigure}
  \begin{subfigure}{0.24\textwidth}
    \includegraphics[width=\textwidth]{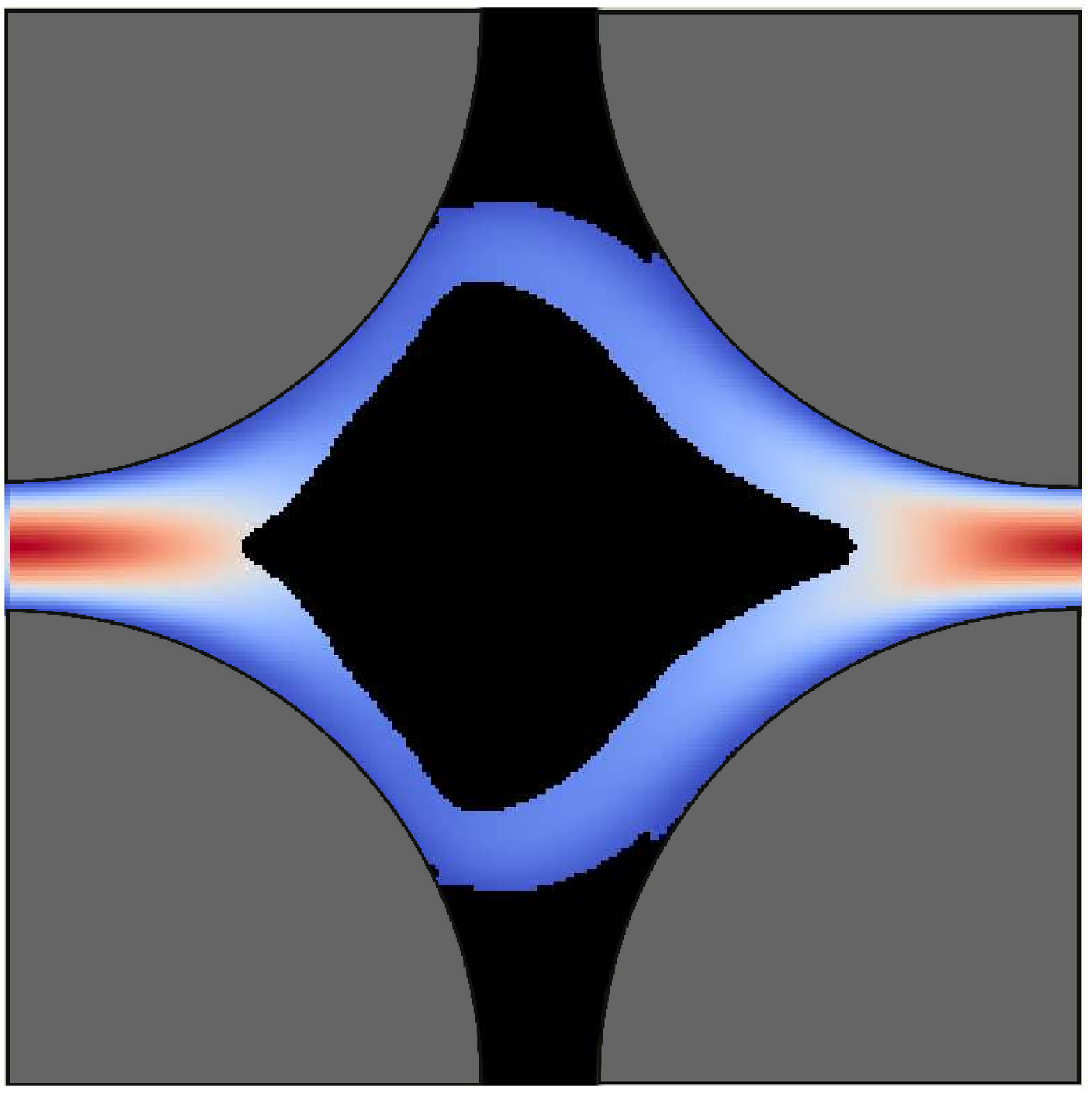}
    \caption{}
    \label{fig:sR_4_100}
  \end{subfigure}
  \caption{Contour of the streamwise velocity component and solid region for different Bingham numbers $Bi$: a) $0.1$, b) $1$, c) $10$, d) $100$. Red corresponds to the maximum velocity and blue to the minimum velocity, while the black area represents the unyielded region. The range of velocity is: a) [-0.008:5.13], b) [-0.003,5.15], c) [-0.012:5.23], d) [-0.014,5.25]. The Reynolds number $Re = 0.1$ for all cases displayed.}
  \label{fig:sR_4}
\end{figure*}
\begin{figure*}[h]
  \centering
  \begin{subfigure}{0.24\textwidth}
    \includegraphics[width=\textwidth]{solid_v1Bi100.eps}
    \caption{}
    \label{fig:sB_1_100}
  \end{subfigure}
  \begin{subfigure}{0.24\textwidth}
    \includegraphics[width=\textwidth]{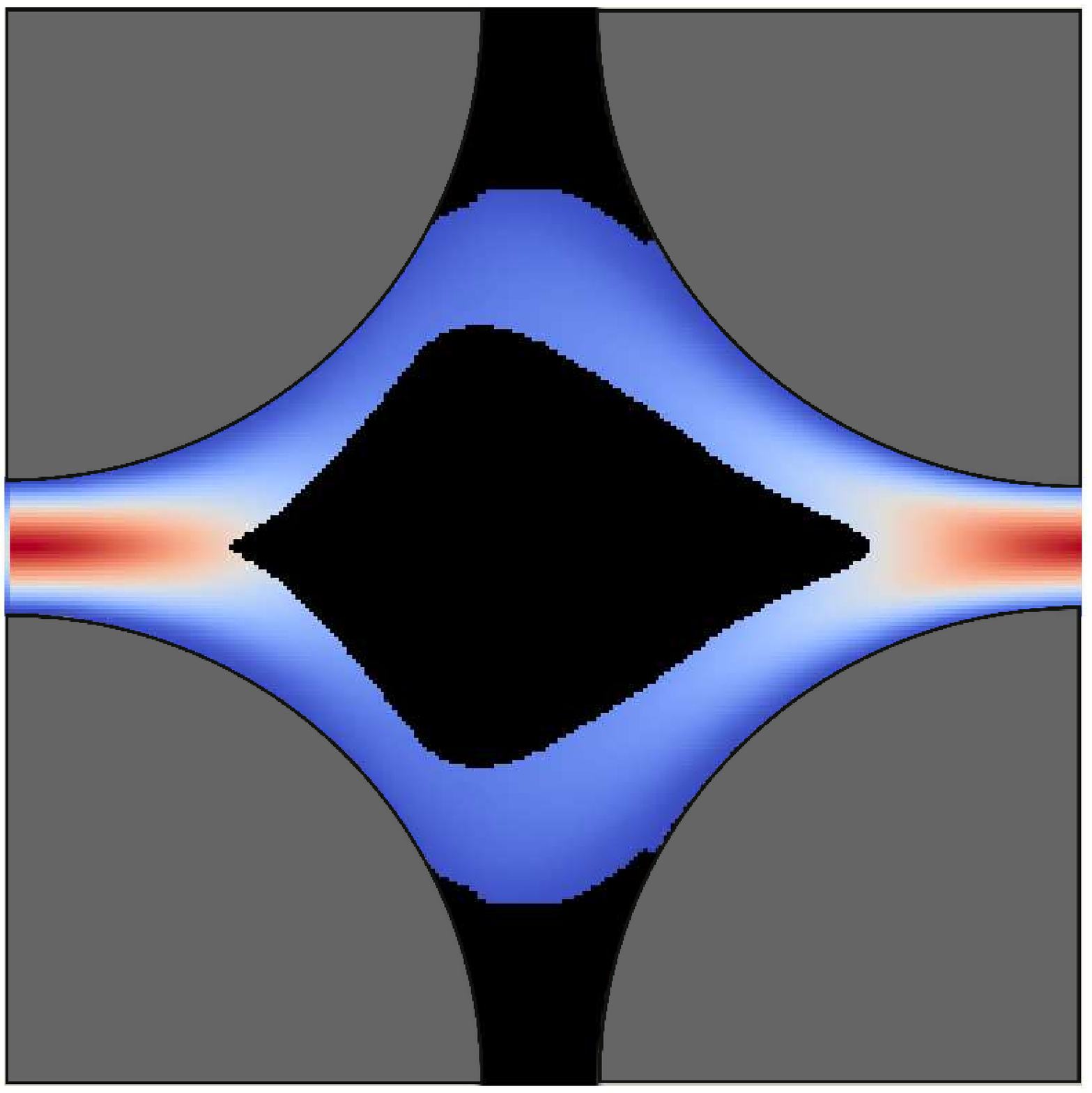}
    \caption{}
    \label{fig:sB_4_100}
  \end{subfigure}
  \begin{subfigure}{0.24\textwidth}
    \includegraphics[width=\textwidth]{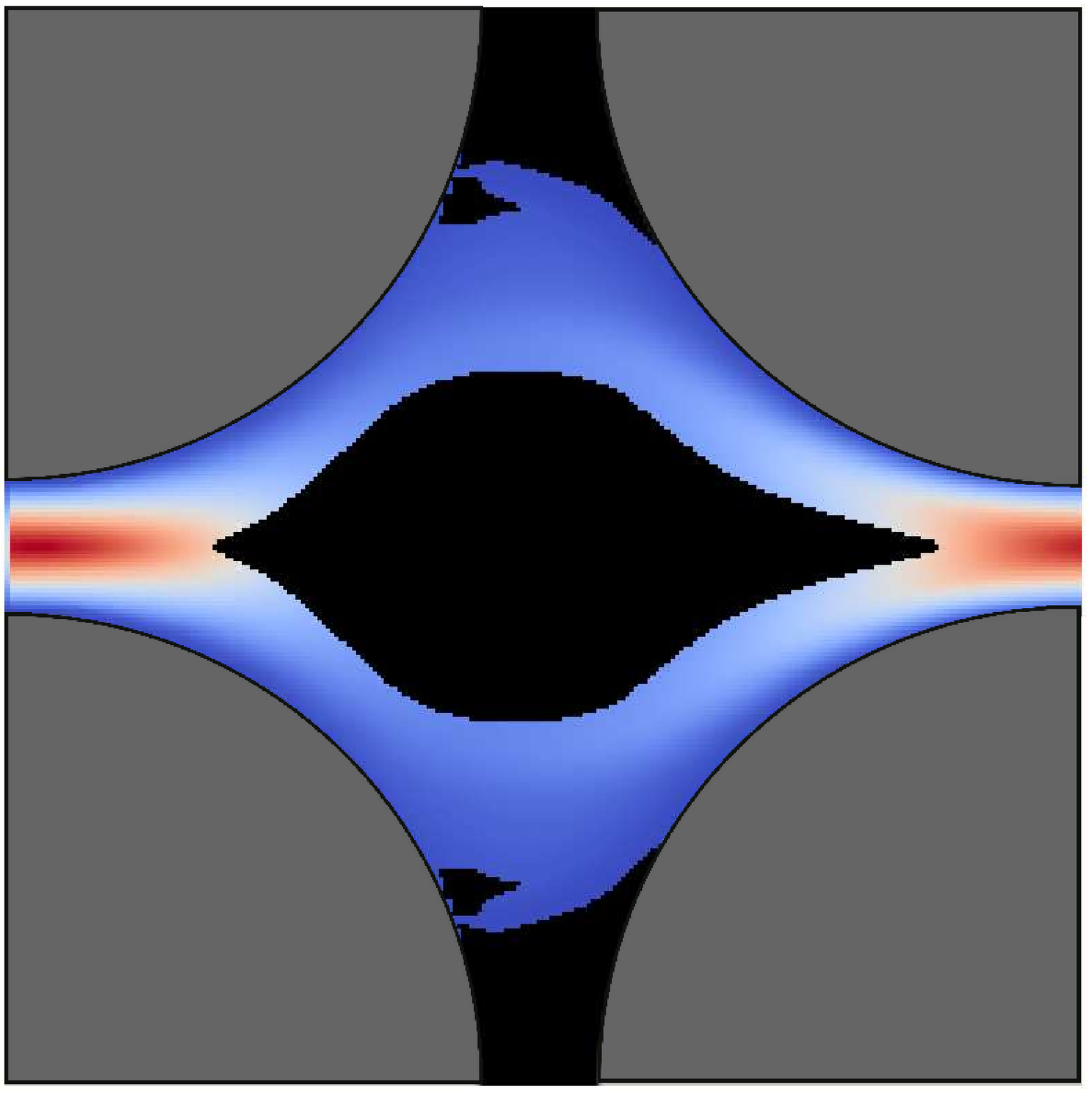}
    \caption{}
    \label{fig:sB_8_100}
  \end{subfigure}
  \begin{subfigure}{0.24\textwidth}
    \includegraphics[width=\textwidth]{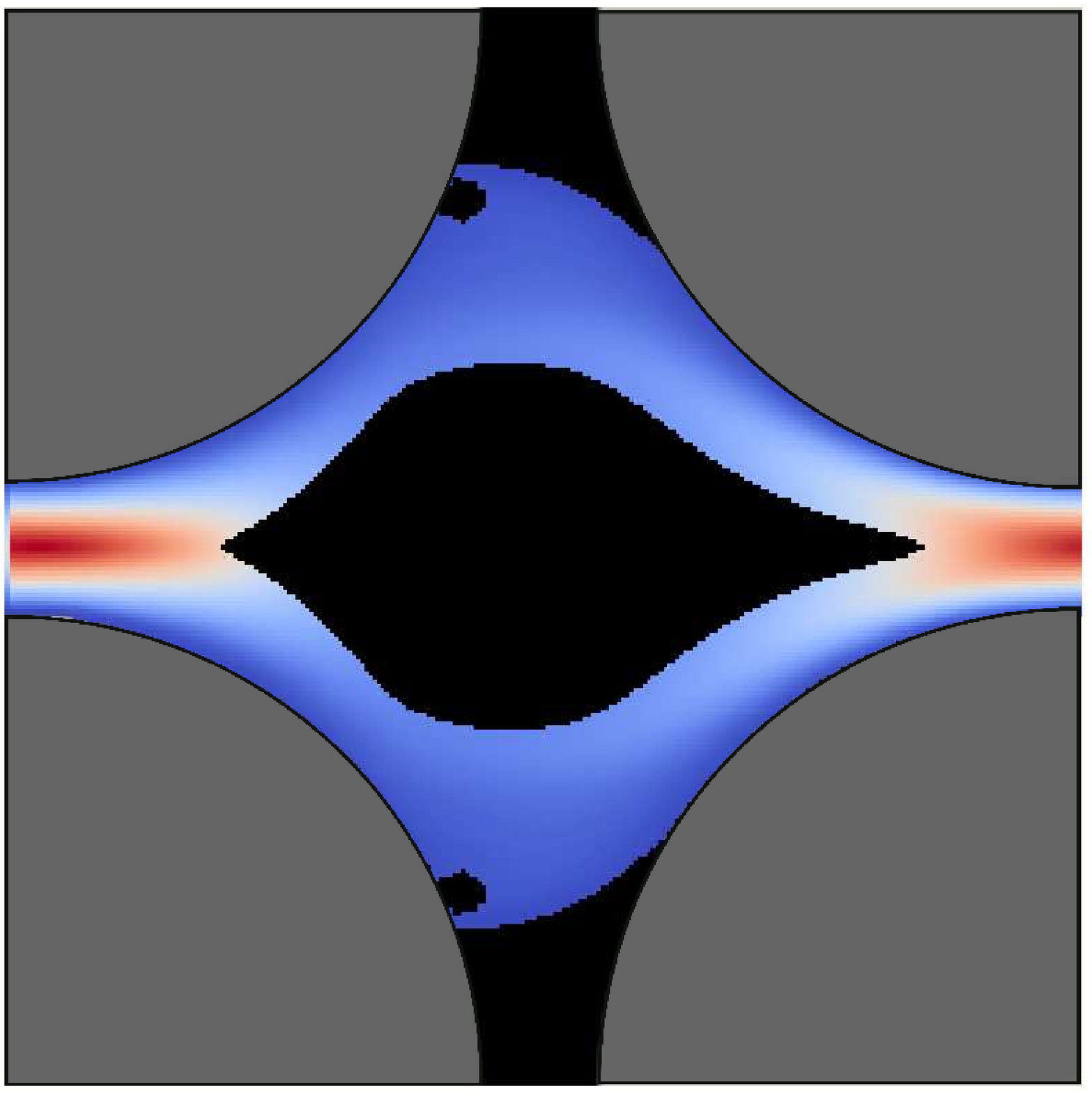}
    \caption{}
    \label{fig:sB_16_100}
  \end{subfigure}
  \caption{Contour of the streamwise velocity component and solid region for different Reynolds numbers $Re$: a) $0.1$, b) $0.4$, c) $0.8$, d) $1.6$. Red corresponds to the maximum velocity and blue to the minimum velocity, while the black area represents the unyielded region. The range of velocity is: a) [-0.014:5.25], b) [-0.0125:5.325], c) [-0.0067:5.5], d) [-0.005:5.57]. The data pertain the simulations at fixed Bingham number $Bi  = 100$.}
  \label{fig:sB_100}
\end{figure*}

Due to the unsteady nature of the flow and the symmetry of the cell, we average all quantities over time and between the two halves of the domain  with respect to the horizontal axis to double the samples for the statistics.} The mean flow and unyielded regions are shown in figure \ref{fig:sR_4} at a fixed Reynolds number ($Re=0.1$) for all the considered Bingham numbers, and in figure \ref{fig:sB_100} for all the Reynolds number at fixed Bingham number ($Bi=100$). At low Bingham number (figure \ref{fig:sR_4_01}) only the small regions located close to the top and bottom boundaries present values of the stresses lower than the yield stress $\tau_0$, where the fluid behaves as a viscoelastic solid (black region in figure). As the Bingham number increases, these two unyielded area becomes larger and larger, untill they completely fill the gap between the cylinders at the top and bottom of the domain (figures \ref{fig:sR_4_1}-\ref{fig:sR_4_100}). A second, disconnected, solid region is generated along the horizontal centerline; this quickly grows when increasing the Bingham number (figure \ref{fig:sR_4_10} and \ref{fig:sR_4_100}). For a fixed Bingham number (figure \ref{fig:sB_100})  the stress increases as the flow rate increases, thus the stresses overcome the yield stress value over a larger portion of the domain and the material behave as a viscoelastic fluid. In particular, the central unyielded region becomes thinner and more stretched in the streamwise direction as the Reynolds number is increased, due to the increased streamwise velocity.

To further investigate the shape of the unyielded regions, we analyze the velocity and stress distribution for the case with $Bi = 10$ and $Re = 0.1$ (see figure \ref{fig:iso}). The expansion and the contraction, that the flow undergoes as consequence of the interaction with the porous medium walls, are responsible for the sharp head and tail of the solid region; the shape of the unyielded region follows the curvature of the four quarters of sphere located in the corners of the domain. The center of this solid region, instead, is influenced by the unyielded region located between the gap in the top and bottom part of the domain. Both stress components, $\tau_{xx}$ and $\tau_{xy}$, follow the shape of the unyielded region in the leading and trailing edges whereas the hollows shape in the middle seems to be a consequence of the antisymmetry of $\tau_{xy}$ with respect to central axes.

\begin{figure*}[]
  \centering
  \begin{subfigure}{0.22\textwidth}
    \includegraphics[width=\textwidth]{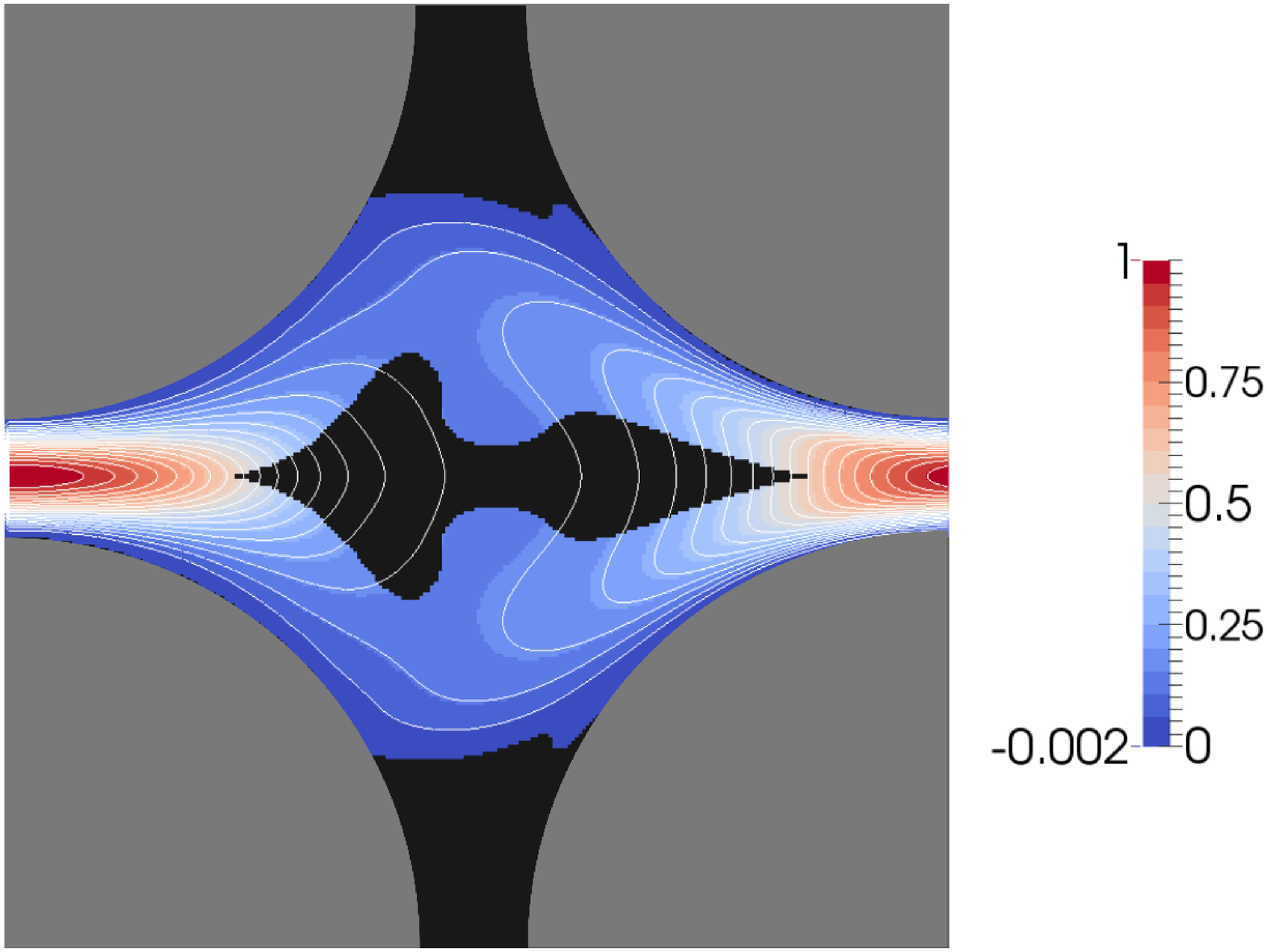}
    \caption{}
    \label{fig:isoV}
  \end{subfigure}
  \begin{subfigure}{0.22\textwidth}
    \includegraphics[width=\textwidth]{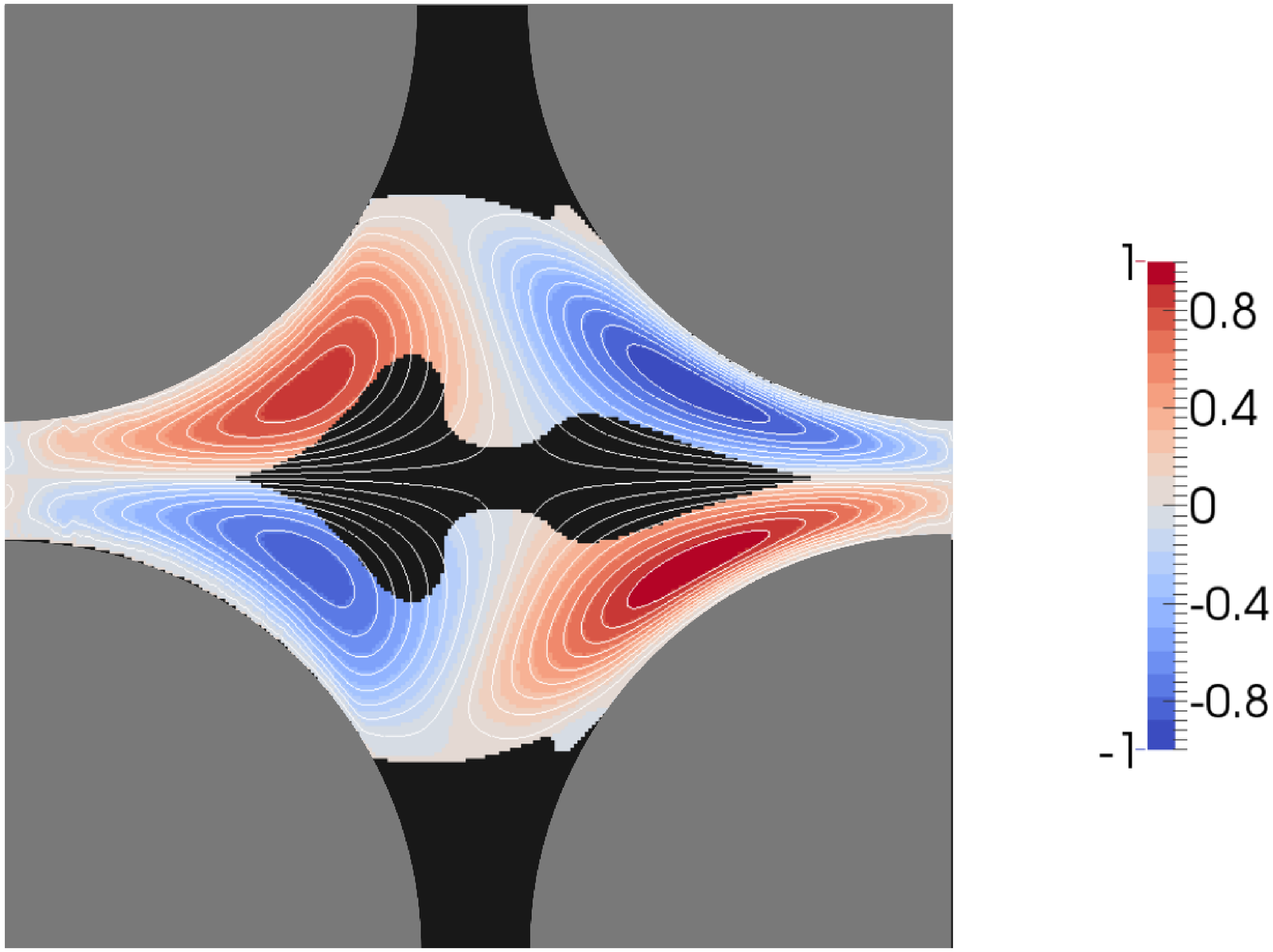}
    \caption{}
    \label{fig:isoW}
  \end{subfigure}
  \begin{subfigure}{0.22\textwidth}
    \includegraphics[width=\textwidth]{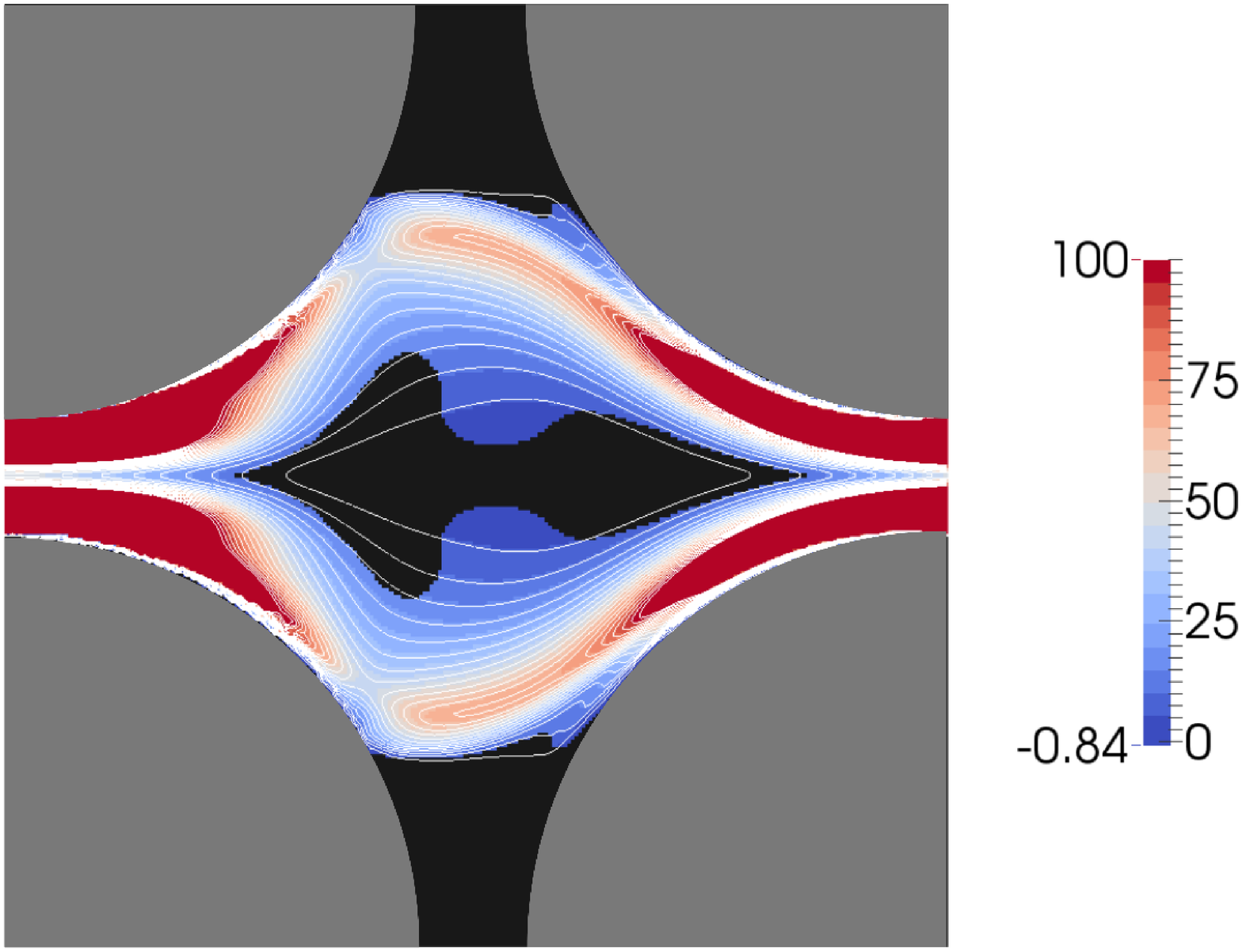}
    \caption{}
    \label{fig:isoC22}
  \end{subfigure}
  \begin{subfigure}{0.22\textwidth}
    \includegraphics[width=\textwidth]{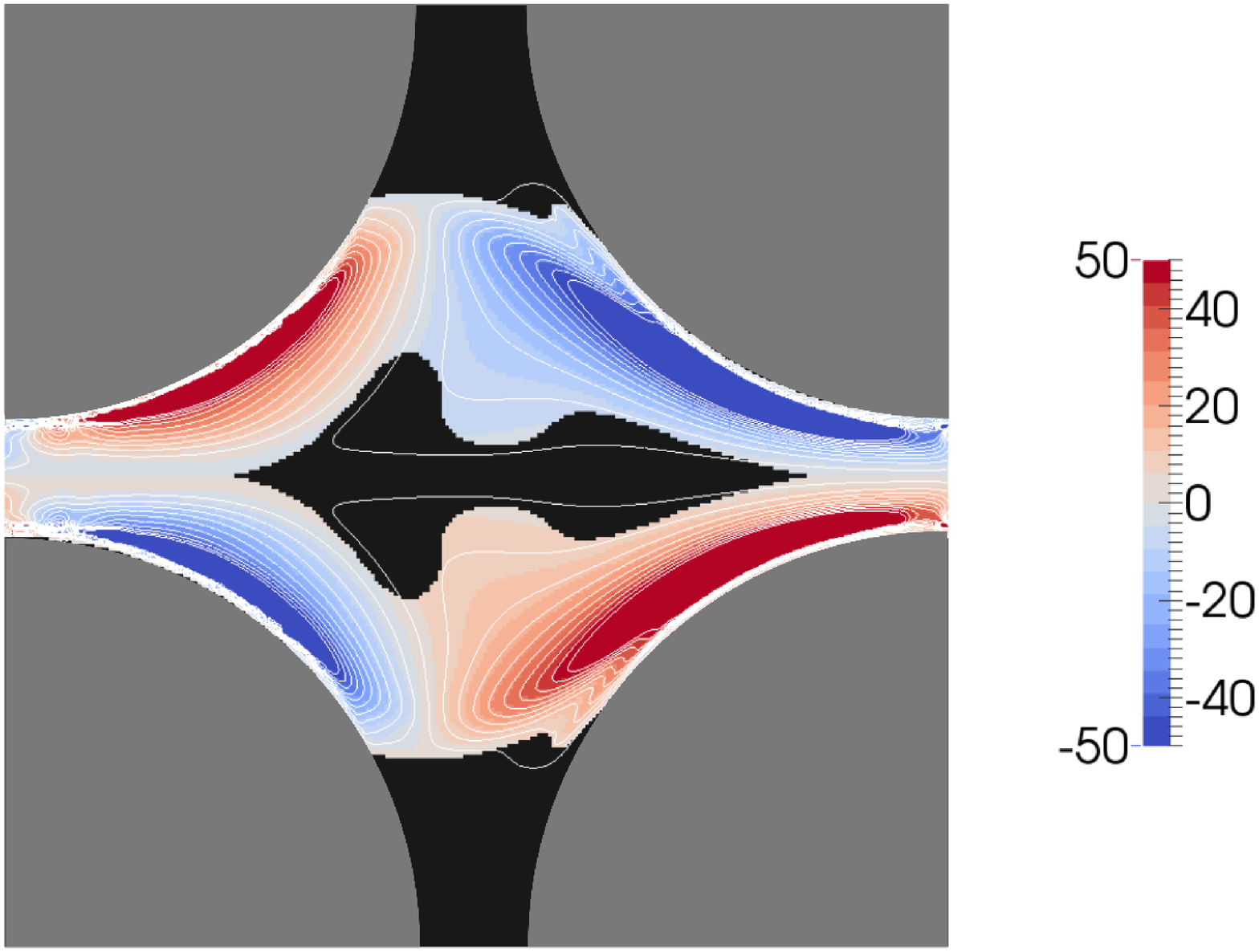}
    \caption{}
    \label{fig:isoC23}
  \end{subfigure}
  \caption{Contour and isolines of: a) streamwise velocity (normalized with the maximum value); b) vertical velocity (normalized with the maximum value); c) stress component $\tau_{xx}$; d) stress component $\tau_{xy}$. The stresses are normalized with the characteristic viscous stress $\mu U / r$. The Bingham number is $Bi = 10$ and the Reynolds nuumber is $Re = 0.1$.}
  \label{fig:iso}
\end{figure*}

In order to quantify the effect of the Reynolds and Bingham numbers on the plastic behavior of the flow, we compute the mean volume of the unyielded region $V_s$ for all the cases investigated and report those in figure \ref{fig:int_solid} as a function of the Reynolds number for different Bingham numbers. 
We observe that the total volume of solid material is nearly unaffected by the Reynolds number, with a decrease of only about $5\%$ between the lowest and highest Reynolds numbers studied, $Re = 0.1$ and $1.6$ respectively. Conversely, $V_s$ is strongly dependent on the Bingham number; in particular,
few percent of the volume are unyielded for $Bi = 0.1$ while more than $70\%$ for $Bi = 100$. 
The vertical bars reported in the figure represent the root mean square (r.m.s.) of the volume integral and quantify the fluctuations of the unyielded region. These grow with the Bingham number reaching a magnitude of about $10\%$ of the total volume for the highest $Bi$ considered. 

\begin{figure}[]
  \centering
  \includegraphics[width=0.45\textwidth]{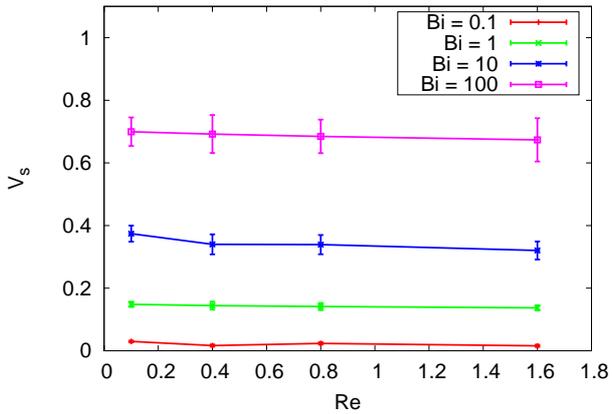}
  \caption{Volume integral of the mean unyielded region, $V_s$, as a function of the Reynolds number, $Re$, for different Bingham numbers, $Bi$. The vertical bars represent the r.m.s.~of the volume integral.}
  \label{fig:int_solid}
\end{figure}

\subsection{Velocity profiles and flow topology}

We continue our analysis by studying the velocity profiles. Figure \ref{fig:stvel} shows the streamwise velocity profiles in the vertical section at $x = L/2$ for all values of  $Re$ and $Bi$ considered. When Bingham is zero, the material is a viscoelastic fluid everywhere and the reduction of the maximum streamwise velocity in the centerline is due to an elastic effect only, while the widening of the velocity profile is a consequence of performing simulations at constant flow rate. Note that, here the Weissenberg number is constant, therefore the elastic effect is the same in all the cases. As the Bingham number increases, the yield stress increases, thus the material located near the centerline behaves as a viscoelastic solid, leading to a further reduction of the maximum velocity and a consequent flattening of the velocity profile in general. Indeed, the velocity profiles tend to become flat; they also exhibit two peaks in the fluid region between the unyielded regions (see figures  \ref{fig:sR_4_100} and \ref{fig:sB_100}). When the Reynolds number is increased, these peaks move towards the centerline as a consequence of the unyielded region at the center becoming thinner (figure \ref{fig:sB_100}). Finally, in the two narrow gaps between the cylinders close to the top and bottom boundaries, the streamwise velocity is very small attaining both positive and negative values.

\begin{figure*}[h]
  \centering
  \begin{subfigure}{0.45\textwidth}
    \includegraphics[width=\textwidth]{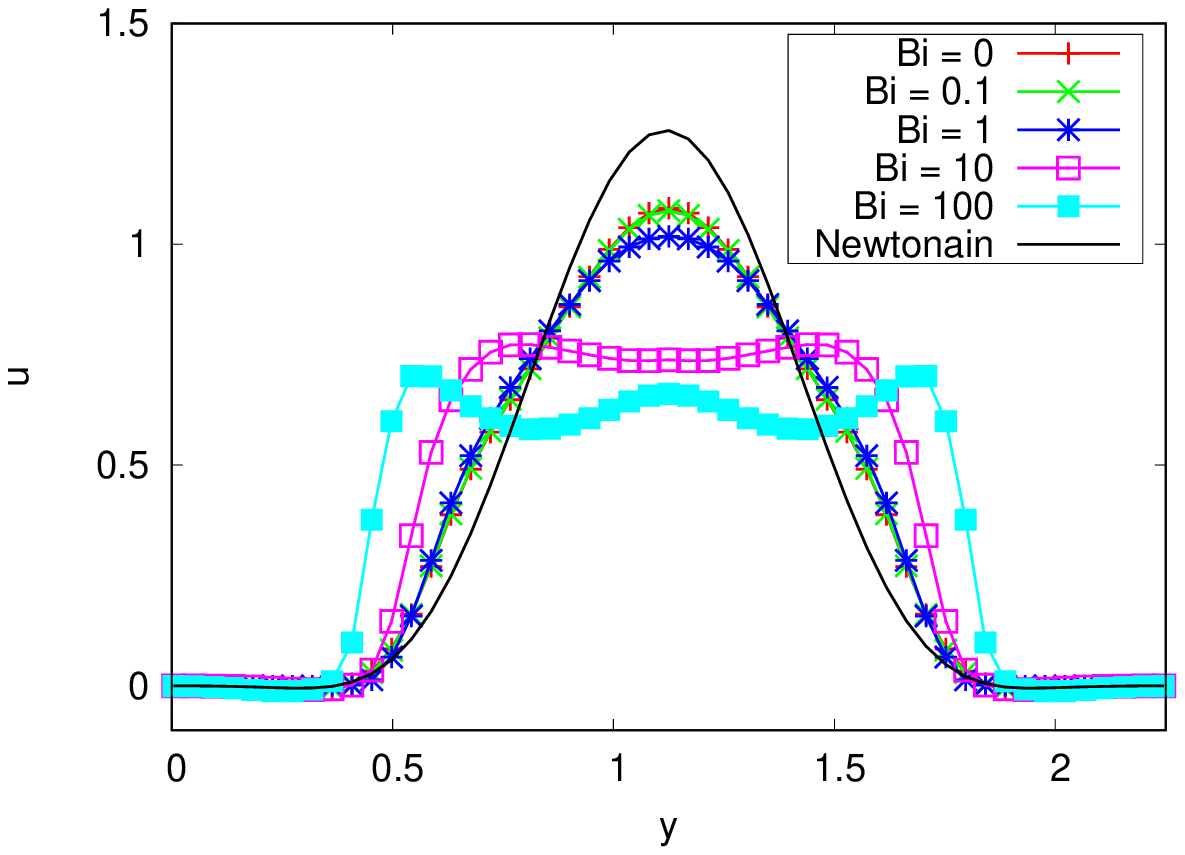}
    \caption{}
    \label{fig:stvel1}
  \end{subfigure}
  \begin{subfigure}{0.45\textwidth}
    \includegraphics[width=\textwidth]{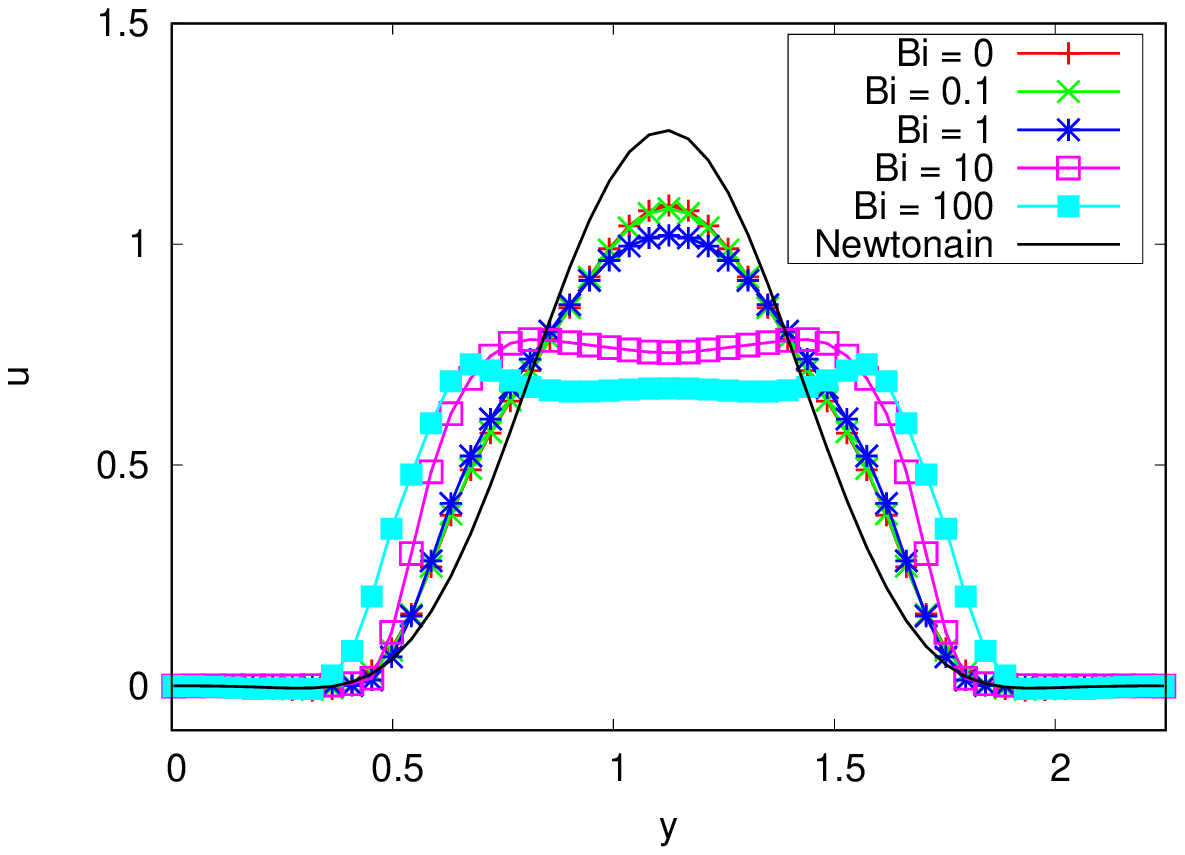}
    \caption{}
    \label{fig:stvel4}
  \end{subfigure}
  \begin{subfigure}{0.45\textwidth}
    \includegraphics[width=\textwidth]{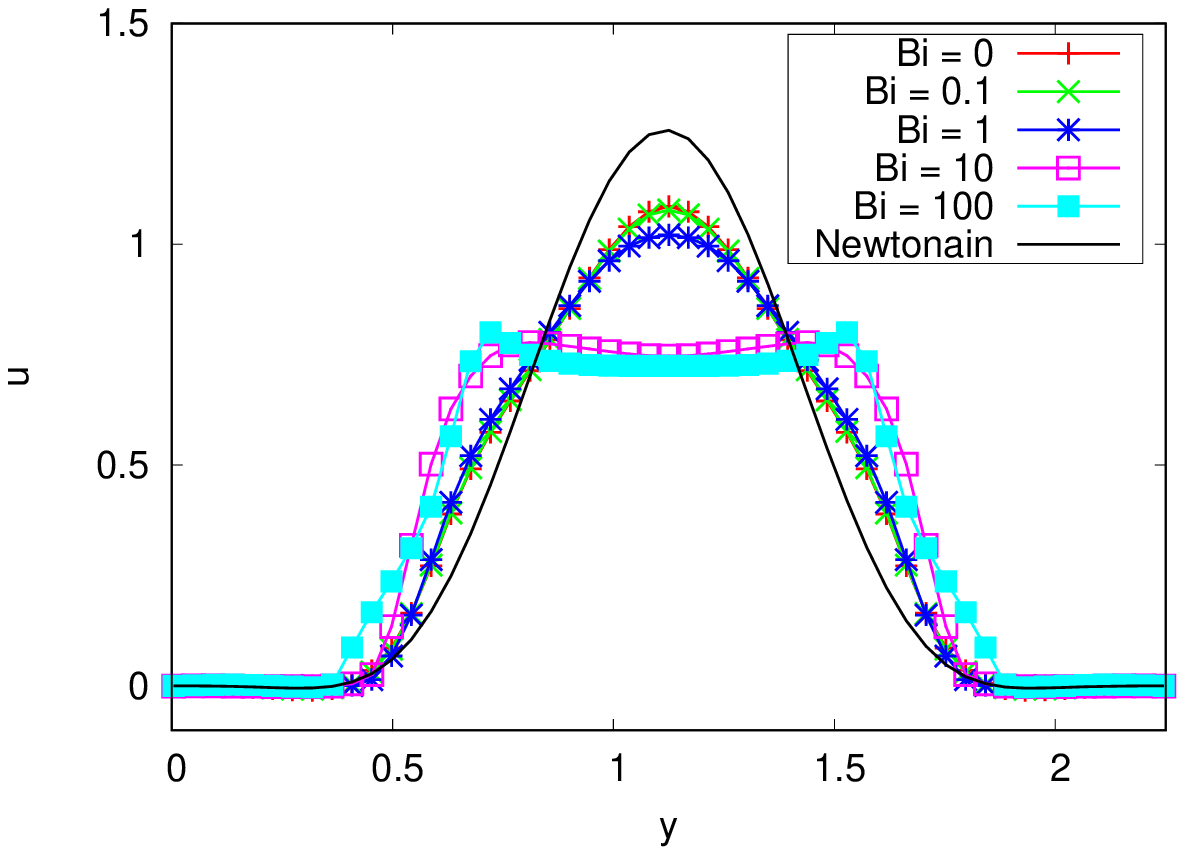}
    \caption{}
    \label{fig:stvel8}
  \end{subfigure}
  \begin{subfigure}{0.45\textwidth}
    \includegraphics[width=\textwidth]{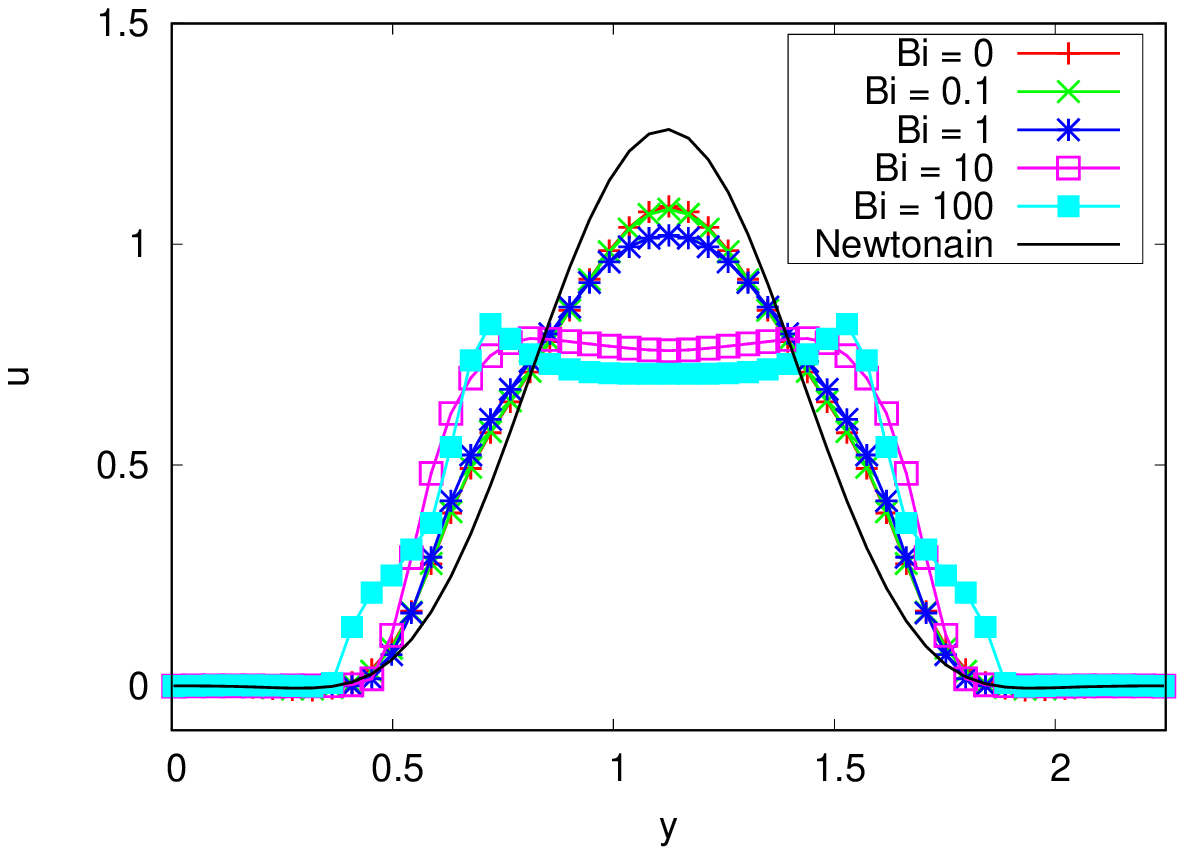}
    \caption{}
    \label{fig:stvel16}
  \end{subfigure}
  \caption{Profile of the streamwise velocity component in the vertical section located at $x = L/2$ for different Bingham numbers, $Bi$, and (a) $Re=0.1$, (b) $0.4$, (c) $0.8$ and (d) $1.6$. The velocity is normalized with the bulk velocity.}
  \label{fig:stvel}
\end{figure*}

Following the analysis by De et al.~\cite{De2017}, we compute the flow topology parameter defined as
\begin{equation} \label{eq:q}
  Q = \frac{D^2 - \Omega ^2}{D^2 + \Omega^2},
\end{equation}
where $D^2 = (\bm{D}:\bm{D})$ and $\Omega^2 = (\bm{\Omega}:\bm{\Omega})$, $\bm{\Omega}$ being the rate of rotation tensor, i.e., $\bm{\Omega} = (\nabla \bm{u}^T - \nabla \bm{u})/2$. When $Q = -1$ the flow is purely rotational, whereas regions with $Q = 0$ represent pure shear flow and those with $Q = 1$ elongational flow. 
The distribution of the flow topology parameter for $Re = 0.1$ and different Bingham numbers is reported in figure \ref{fig:flowtop_v1}. All the curves exhibit a dominant peak in correspondence to $Q = 0$, suggesting that the flow is mostly a shear flow. For the Newtonian case (black curve), the right tail of the curve drops to zero for $Q = 1$ and displays moderate values for $Q$ between $0$ and $0.5$. On the contrary, all the EVP curves exhibit nonzero values for $Q=1$ indicating the presence of purely elongational flow, 
values lower than the Newtonian fluid for $Q$ between $0$ and $0.5$ and, for high Bingham numbers, also negative values of $Q$.
Note that  these distributions of the topology parameter are found to weakly change when increasing the Reynolds number over the range considered in this study and data pertaining different $Re$ are therefore not reported here. 

Further insight can be gained by showing the histogram of $Q$ separately in the yielded and unyielded regions, as reported in figure \ref{fig:flowtop_part} for the flow with $Re = 0.1$ and $Bi = 100$. The data reveal that the yielded part of the flow is the one mostly responsible for the shear flow behavior, whereas shear and elongation flows are almost equally distributed in the unyielded regions.  Similar feature is also observed in flows of ideal visco-plastic fluids in anfractuous configurations \cite{Maleki2015, Roustaei2016}.

\begin{figure}[t]
  \centering
  \begin{subfigure}{0.49\textwidth}
    \includegraphics[width=\textwidth]{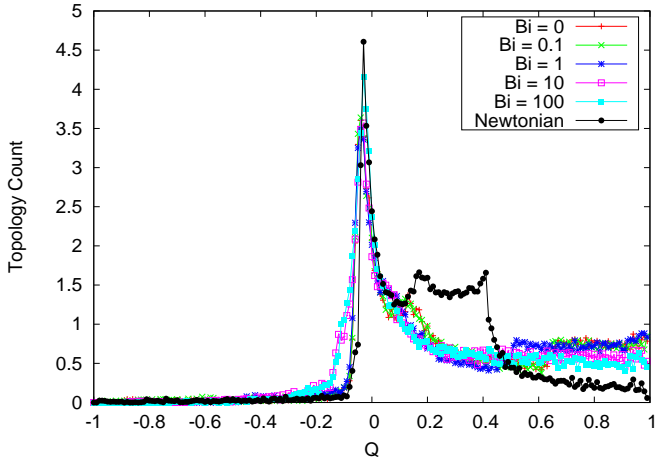}
    \caption{}
    \label{fig:flowtop_v1}
  \end{subfigure}
  \begin{subfigure}{0.49\textwidth}
    \includegraphics[width=\textwidth]{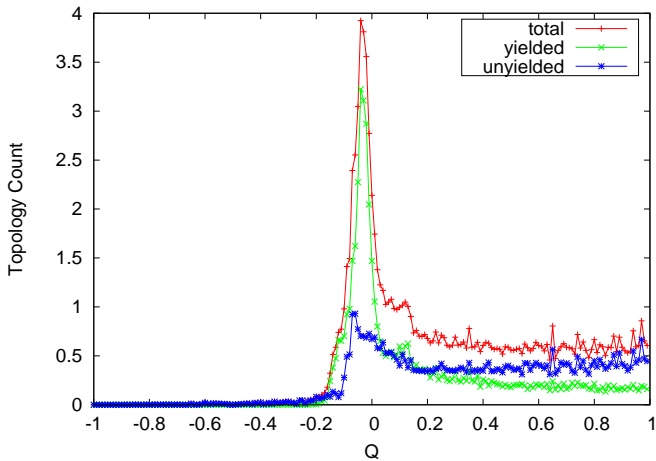}
    \caption{}
    \label{fig:flowtop_part}
  \end{subfigure}
  \caption{Histogram of the flow topology parameter $Q$ defined in equation \ref{eq:q}: a) results for $Re = 0.1$ and different $Bi$; b) contribution of the yield and unyielded region to the histogram for $Re = 0.1$ and $Bi = 0.1$. The curves in (a) and the curve showing the total distribution in (b) are normalized such that the underlying area is equal to 1.}
  \label{fig:flowtop}
\end{figure}

\subsection{Pressure drop and effective permeability}

At last, we analyse the macroscopic behaviour of the EVP flow in a porous medium. In particular, we focus on the
effect of the Reynolds and Bingham numbers on the pressure gradient needed to drive the flow, i.e., to the pressure drop across the domain. 
As already mentioned, all our simulations are performed at a constant flow rate and we compute the instantaneous value of the streamwise pressure gradient required to provide the desired flow rate at each time step. When the fluid is Newtonian, the relation between the mean pressure gradient $\Delta p / L$ and the mean velocity in a porous media is given by the Darcy law
\begin{equation}
  U = - \frac{K}{\mu \varepsilon} \frac{\Delta p}{L},
  \label{eqn:Darcy}
\end{equation}
where $U$ is the mean flow velocity, sometimes called Darcy velocity, $\varepsilon$ is the porosity of the medium and $K$ its permeability. For convenience, we rewrite the previous relation in a non-dimensional form as
\begin{equation}
  \frac{\Delta p^*}{L^*} = \frac{\varepsilon}{\Sigma^2 Re},
  \label{eqn:Darcynondim}
\end{equation}
where $p^*$ is the non-dimensional pressure equal to $p / \rho U^2$, $L^*=L/r$ and $\Sigma$ is the non-dimensional permeability defined as $\Sigma=\sqrt{K}/r$ \cite{rosti_cortelezzi_quadrio_2015a}. For an EVP fluid flowing through a porous media the pressure gradient is in general a function of inertia ($Re$), elasticity ($Wi$), plasticity ($Bi$) and geometry ($\varepsilon$ and other parameters defining the specific configuration of the porous medium), i.e., $\Delta p^*/L^* = \mathcal{F}(Re,$ $We, Bi, \varepsilon, \ldots)$. For now, the Weissenberg number, the porosity $\varepsilon$ and the geometry of the porous medium are kept constant, thus the pressure gradient will vary only with the Reynolds and Bingham numbers, e.g., the other dependencies are dropped. In the next subsection the effect of the Weissenberg number will be examinated.

Figure \ref{fig:pressuredrop} shows the mean pressure gradient as a function of the Reynolds number for different values of the Bingham number (panel a) and as function of the Bingham number for different values of the Reynolds number (panel b). The pressure drop decreases with the Reynolds number whereas it increases as a non-linear function with the Bingham number. In order to derive an expression for $\mathcal{F}$ as close as possible to a Darcy-type law, we assume the function $\mathcal{F}$ to be the product of two terms, one depending on the Reynolds number and the other on the Bingham number. By fitting the data of our numerical simulations with a polynomial expression, we find the following expression to properly describe the relation between pressure drop and the EVP flow
\begin{equation}
  \frac{\Delta p^*}{L^*} = \frac{5.494 Bi^{0.561} + 91.042}{Re},
  \label{eqn:fit}
\end{equation}
which is shown in figure \ref{fig:fit} with the black solid lines. The relation \eqref{eqn:fit} provides an accurate prediction for all the points pertaining our simulations, except perhaps the case at high $Bi$ and $Re$ which slightly deviates from the fitting.
\begin{figure}[h]
  \centering
  \begin{subfigure}{0.49\textwidth}
    \includegraphics[width=\textwidth]{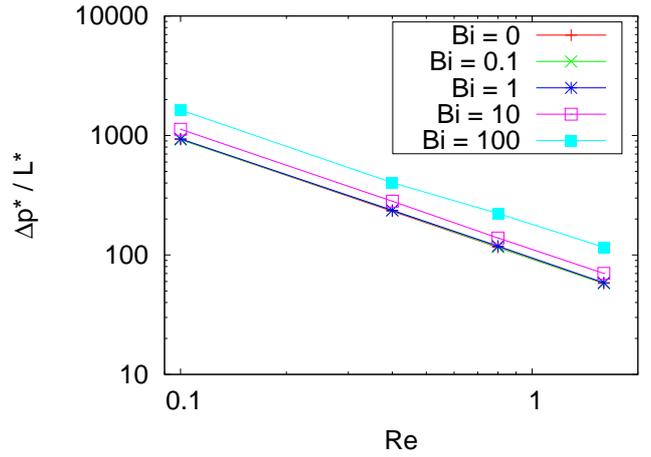}
    \caption{}
    \label{fig:Redpdx}
  \end{subfigure}
  \begin{subfigure}{0.49\textwidth}
    \includegraphics[width=\textwidth]{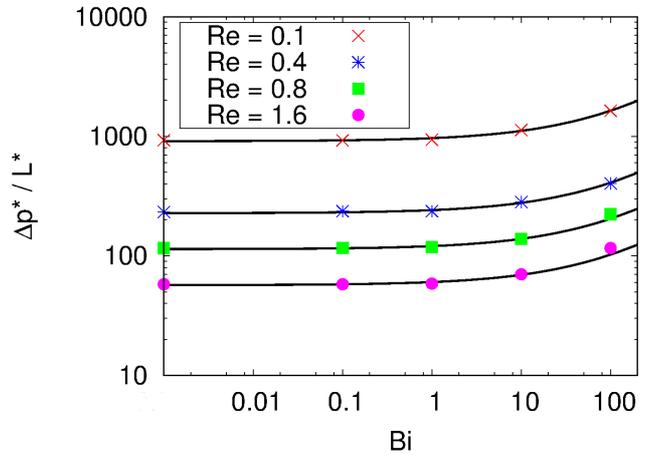}
    \caption{}
    \label{fig:fit}
  \end{subfigure}
  \caption{Non-dimensional mean pressure gradient as a function of (a) the Reynolds number and  (b) of the Bingham number. The black lines in panel b show the fit obtained with equation \eqref{eqn:fit}. Note that, the leftmost points in panel b correspond to the purely viscoelastic case with $Bi=0$.}
  \label{fig:pressuredrop}
\end{figure}

It is worth noticing that the coefficients in equation \eqref{eqn:fit} are valid only for the specific set of parameters chosen in this study, as we do not consider the dependency on the geometry and the Weissenberg number. As for the fit proposed in Ref.~\cite{Chevalier2013} for pure yield-stress fluids, the pressure drop increases with the yield stress, i.e., with the Bingham number, although with the following differences. 
In fact, the experiments in Ref.~\cite{Chevalier2013} deal with a creeping flow, \emph{i.e.} zero Reynolds number, of a pure yield stress fluid in a 3D configuration. The authors proposed a fit assuming a Herschel-Bulkley model plus an additional term, corresponding to the minimum pressure drop required to start the flow. With these hypotheses, they found a linear relation between the pressure drop and the yield stress and obtained fitting coefficients from the experimental data. In the present study, we take into account both inertia end elasticity, which leads to a different Darcy-type law. In particular, we find the pressure drop to depend on the square root of the Bingham number and to be inversely proportional to the Reynolds number.

We also note that other authors have shown a strong dependency of the pressure drop on the geometry; indeed, De et al \cite{De2017} document a significant difference between a symmetric and an asymmetric array of cylinders in terms of pressure drop, whereas Rouestaei et al \cite{Roustaei2016} state that the approximation error given by a Darcy-type law to compute the pressure drop in fracture flows strongly increases for large heights of the fracture.

It may be instructive to write equation \eqref{eqn:fit} in a form similar to the one proposed in Ref.~\cite{Chevalier2013}:
\begin{equation}
  \frac{\Delta p}{L} = C_1 \left( \frac{\tau_0 r}{\mu U} \right)^{C_2} \frac{\mu U}{r^2} + C_3 \frac{\mu U}{r^2}.
  \label{eqn:dpcoeff}
\end{equation}
As already said before, the coefficients $C_1$, $C_2$ and $C_3$ depend, in general, on the geometry and the Weissenberg number. This result, along with the recent study in Ref.~\cite{Roustaei2016}, strongly suggests that it may  not be possible to derive a simple general law for non-Newtonian flow in porous media.

Equation \eqref{eqn:fit} reduces to a form similar to equation \eqref{eqn:Darcynondim} for $Bi = 0$, but with a different pressure drop than for the Newtonian case ($120/Re$), due to the elasticity effects. In particular, we find that the permeability in the viscoelastic case is $25\%$ higher than in the Newtonian case, in qualitative agreement with the results in \cite{De2017}. Assuming a relation between the pressure drop and the flow rate of the same form as in equation \eqref{eqn:Darcy}, we can define an apparent permeability by computing the ratio of the pressure gradients of the Newtonian and non-Newtonian fluids and  obtain the following relation
\begin{equation}
  \kappa_{app} = \frac{\kappa_{EVP}}{\kappa_{N}} = \frac{(\Delta p)_{N}}{(\Delta p)_{EVP}}.
  \label{eqn:apparent}
\end{equation}
Figure \ref{fig:apparent} shows the apparent permeability as a function of the Bingham number for all the considered Reynolds numbers. At low Bingham, i.e., $Bi < 10$, the elastic effects are dominant and the apparent permeability is greater than 1, indicating a reduction of the pressure gradient required to drive the flow. On the contrary, for high Bingham numbers, the apparent permeability decreases below $1$, indicating higher pressure drops. Indeed, at high Bingham numbers, a larger portion of the fluid behave as a viscoelastic solid, hence the velocity is higher in the unyielded regions, corresponding to higher shear rates and rotational motions and thus to higher dissipation, as previously shown in figure \ref{fig:flowtop_v1}. The effect of the Weissenberg number on the apparent viscosity for a viscoelastic flow in porous media is discussed in Ref.~\cite{De2017}. These authors assume  the permeability to be constant and relate the ratio of pressure drop as in \eqref{eqn:apparent} to an effective viscosity. However, we believe that the assumption of constant permeability may not ideal due to the flow nonlinearity
and therefore assume a constant viscosity and consider an apparent permeability instead. Note that, by doing as described in Ref.~\cite{De2017} for $Bi = 0$, we obtain an effective viscosity equal to $0.78$, which is close to the value reported by those authors ($0.73$) although for a different viscoelastic model (FENE-P).

\begin{figure}[h]
  \includegraphics[width=0.5\textwidth]{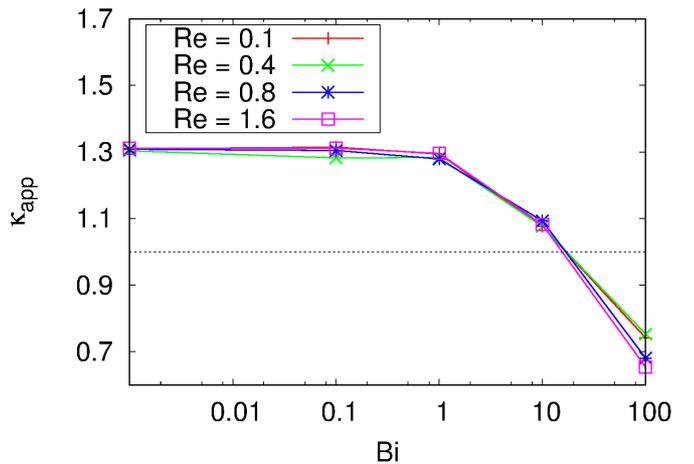}
  \caption{Apparent permeability $\kappa_{app}$ as a function of the Bingham number $Bi$ for different Reynolds numbers $Re$. Note that the leftmost points in the figure correspond to the purely viscoelastic case with $Bi=0$.}
  \label{fig:apparent}
\end{figure}

\subsection{Effect of elasticity}

In this last section, we consider the effect of the elasticity and vary the Weissenberg number for two different Bingham numbers, $Bi = 0$ and $Bi = 10$, and for a fixed value of the Reynolds number, $Re = 1.6$. Since the adopted evolution equation for the stress tensor is based on the Oldroyd-B model, the range of Weissenberg numbers considered here is between $0$ and $0.5$, due to the stability limitation \cite{Saramito2007}.
Figure \ref{fig:Wisolid} shows the unyielded region distribution inside the porous medium for three different values of the Weissenberg number. The volume of the solid region, $V_s$, increases with the Weissenberg number, being 0.298 for $Wi = 0.1$, 0.306 for $Wi = 0.25$ and 0.320 for $Wi = 0.5$. Recalling the sketch in figure \ref{fig:model}, the Weissenberg number is proportional to the relaxation time $\lambda$ which is inversely proportional to the spring stiffnes $\kappa$. Therefore, for higher Weissenberg numbers the elastic deformation in the material is larger and the unyielded region located at the center of the domain can stretch following the expansion and contraction of the flow. Conversely, for low Weissenberg numbers the material is stiffer and exhibits smaller deformation.

\begin{figure}[h]
  \centering
  \begin{subfigure}{0.15\textwidth}
    \includegraphics[width=\textwidth]{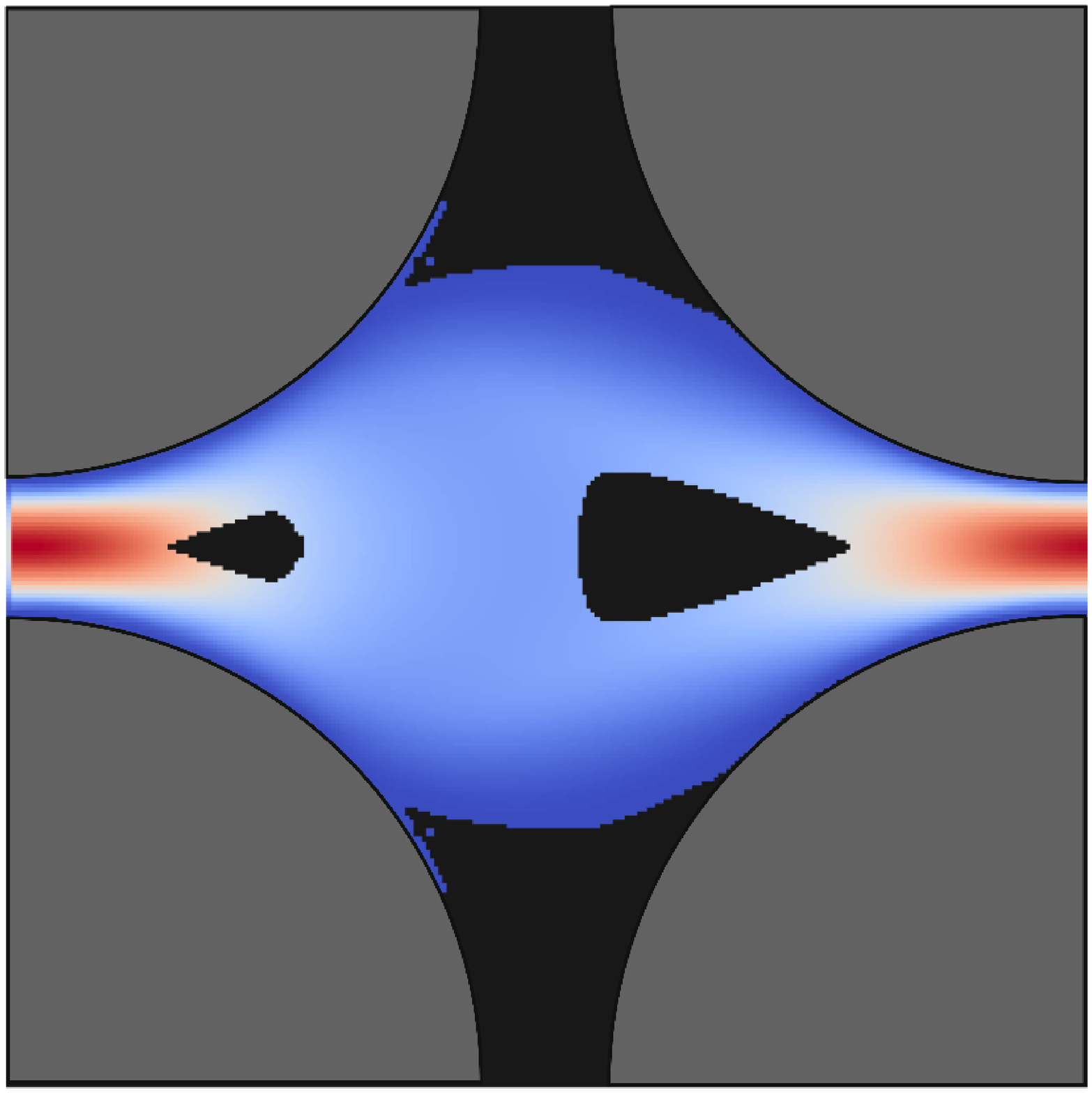}
    \caption{}
    \label{fig:solidWi01}
  \end{subfigure}
  \begin{subfigure}{0.15\textwidth}
    \includegraphics[width=\textwidth]{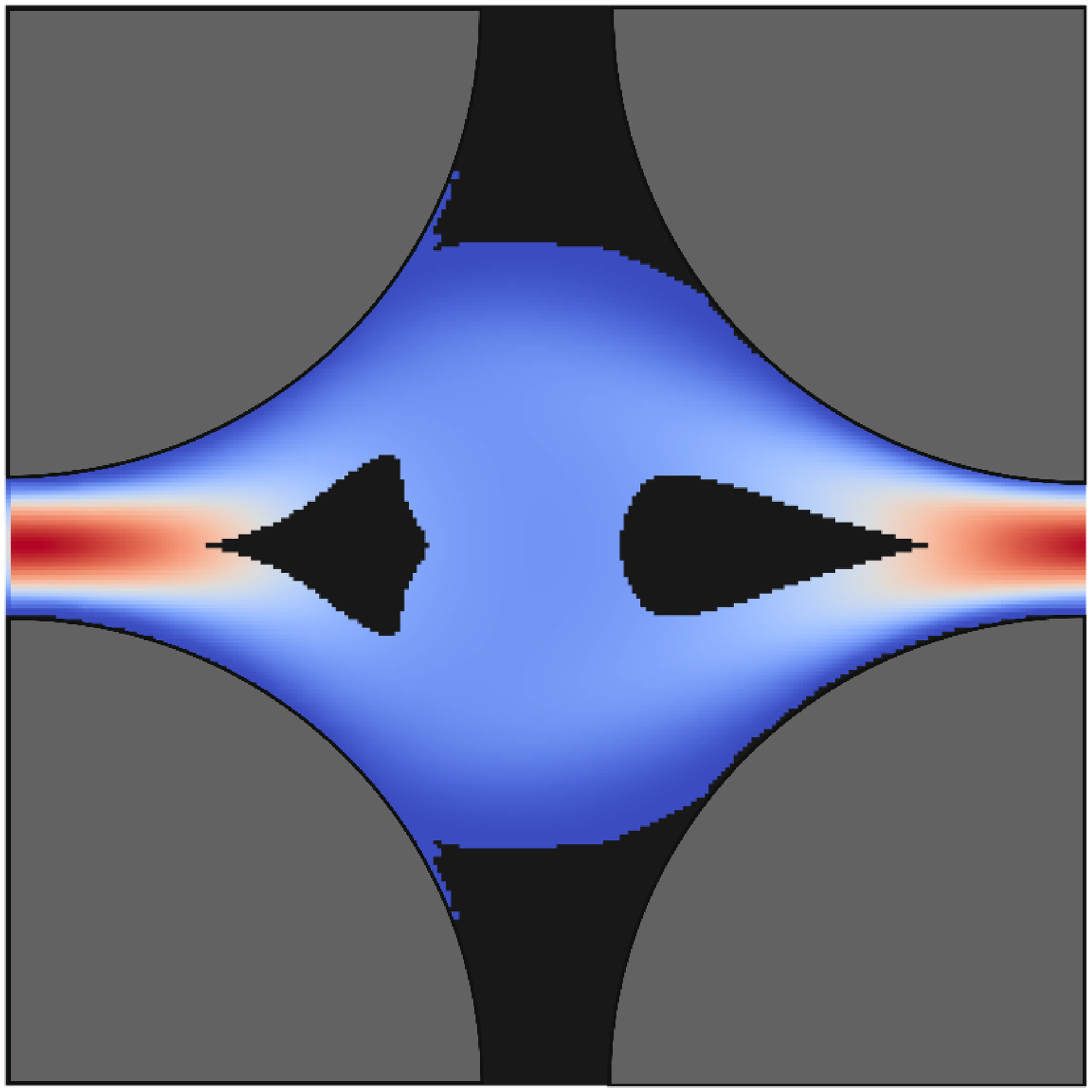}
    \caption{}
    \label{fig:solidWi025}
  \end{subfigure}
  \begin{subfigure}{0.15\textwidth}
    \includegraphics[width=\textwidth]{solid_v1Bi10.eps}
    \caption{}
    \label{fig:solidWi05}
  \end{subfigure}
  \caption{Streamwise velocity contour and unyielded region (solid black) for three different values of the Weissenberg number: a) $Wi = 0.1$; b) $Wi = 0.25$; c) $Wi = 0.5$. For all configurations Bingham number is $Bi = 10$ and Reynolds number is $Re = 1.6$.}
  \label{fig:Wisolid}
\end{figure}

Finally, we analyze the effect of the elasticity on the overall pressure gradient. Figure \ref{fig:Wiapparent} displays the non-dimensional mean pressure gradient as a function of the Weissenberg number for the two different values of the Bingham number examined. For the viscoelastic case, \emph{i.e.} $Bi = 0$, we find a trend similar to that reported in Ref.\ \cite{De2017}, with a minimum of the pressure gradient between $Wi = 0.2$ and $Wi = 0.3$. The difference with this previous work is small, below 5\%, which can be a consequence of the different model for the stress evolution equation.

As regards the correlation between pressure drop and flow rate in equation \ref{eqn:dpcoeff}, we have assumed the coefficients $C_1$, $C_2$ and $C_3$ to be function of the geometry and of the Weissenberg number. Since for $Bi = 0$ the elastic effects must be included $C_3 = C_3(Wi)$. If we vertically translate the curve for $Bi = 10$ such that the point corresponding to $Wi = 0.1$ overlaps with the same point of the curve for $Bi = 0$, we can see that there is a weak combined effect of the Bingham number and of the Weissenberg number on the pressure drop, which makes $C_1$ and $C_2$, or at least one of the two, also function of the Weissenberg number. Additional studies are therefore needed to clarify this point.

\begin{figure}
  \centering
  \includegraphics[width=0.45\textwidth]{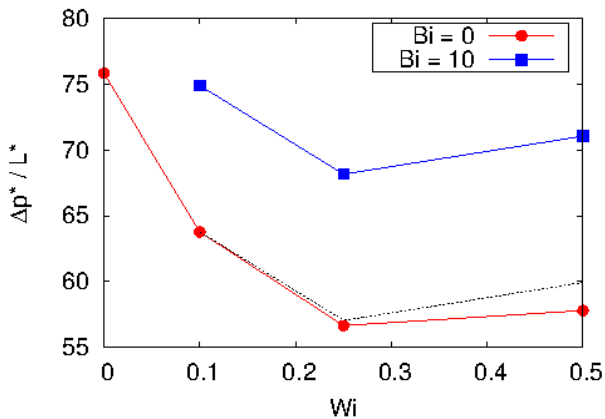}
  \caption{Non-dimensional mean pressure gradient as a function of the Weissenberg number for two different values of the Bingham number: $Bi = 0$ (red dot) and $Bi = 10$ (blue square). The Reynold number is $Re = 1.6$. The dotted curve is the same as $Bi = 10$ but translated vertically in order to overlap with the curve for $Bi = 0$ corresponding to $Wi = 0.1$.}
  \label{fig:Wiapparent}
\end{figure}

\section{Conclusion}
We have performed numerical simulations of the elastoviscoplastic flow through porous media modelled as an array of cylinders, and considered a single periodic cell. The flow is described by the Navier-Stokes equations, and the additional evolution equation for the EVP stress tensor following the model proposed by Saramito \cite{Saramito2007}. 

We find the flow dynamics to be time-dependent, and have thus presented both instantaneous configurations and time-averaged results.
The data show that the volume where the fluid is unyielded strongly increases with the Bingham number, whereas it slowly decreases with the Reynolds number.
The unsteadiness of the flow is measured in terms of the r.m.s of the volume of yielded fluid,  which, in turn, produces oscillations in the pressure gradient; this unsteadiness grows with the Bingham 
number. Due to the unyielded fluid at the center of the domain, the maximum velocity at the centerline decreases and the velocity profile flattens. From the analysis of the flow topology, we show that the flow is mainly a shear flow in the yielded region whereas it is equally elongated and sheared in the unyielded part of the domain.

The analysis of the data allowed us to extract a relation between the pressure drop across the domain and the Reynolds and Bingham numbers. For low Bingham numbers the elastoviscoplastic flow is characterized by an apparent permeability higher than that of a Newtonian flow, corresponding to a smaller pressure drop. On the contrary, the apparent permeability is smaller than  that of a Newtonian flow in a porous media for high Bingham numbers, corresponding to higher pressure drops. No generalization of Darcy law for non-Newtonian flows that is generally valid in many different conditions is available in literature. Results presented in our study and other recently proposed in \cite{Roustaei2016}, seem to confirm that it is not possible to derive a general form of the Darcy law to describe complex non-Newtonian flows through porous media.

The analysis presented here considers a periodic cell of a porous media made of a symmetric array of cylinders. Although this assumption is widely adopted in literature, this configuration is not fully representative of real porous media. A possible extension of this work is therefore the investigation of the flow in more complex geometries,
such as a random distribution of cylinders/spheres. 
Additionally, future studies are needed to deeper investigate the combined effects of elasticity and plasticity on the dynamics of the flow.

\section*{Acknowledgment}
This work was supported by the European Research Council Grant no. ERC-2013-CoG-616186 and TRITOS. We also acknowledge financial support by the Swedish Research Council through grants No. VR 2013-5789, No. VR 2014-5001 and No. VR 2017-76478. S.H. acknowledges financial support by NSF  (Grant No. CBET-1554044-CAREER), NSF-ERC (Grant No. CBET-1554044 Supplementary CAREER) and ACS PRF (Grant No. 55661-DNI9). The authors acknowledge computer time provided by SNIC (Swedish National Infrastructure for Computing).

\section*{References}
\bibliography{biblio}

\end{document}